# A new, very massive modular Liquid Argon Imaging Chamber to detect low energy off-axis neutrinos from the CNGS beam.

## (Project MODULAr)

B. Baibussinov[1], M. Baldo Ceolin[1], G. Battistoni[2], P. Benetti[3], A. Borio[3], E. Calligarich[3], M. Cambiaghi[3], F. Cavanna[4], S. Centro[1], A. G. Cocco[5], R. Dolfini[3], A. Gigli Berzolari[3], C. Farnese[1], A. Fava[1], A. Ferrari[2], G. Fiorillo[5], D. Gibin[1], A. Guglielmi[1], G. Mannocchi[6], F. Mauri[3], A. Menegolli[3], G. Meng[1], C. Montanari[3], O. Palamara[4], L. Periale[6], A. Piazzoli[3], P. Picchi[6], F. Pietropaolo[1], A. Rappoldi[3], G.L. Raselli[3], C. Rubbia[A][4], P.Sala[2], G. Satta[6], F. Varanini[1], S. Ventura[1], C. Vignoli[3]

*[1]Dipartimento di Fisica e INFN, Università di Padova, via Marzolo 8, I-35131*
*[2]Dipartimento di Fisica e INFN, Università di Milano, via Celoria 2, I-20123*
*[3]Dipartimento di Fisica Nucleare, Teorica e INFN, Università di Pavia, via Bassi 6, I-27100*
*[4]Laboratori Nazionali del Gran Sasso dell'INFN, Assergi (AQ), Italy*
*[5]Dipartimento di Scienze Fisiche, INFN and University Federico II, Napoli, Italy*
*[6]Laboratori Nazionali di Frascati (INFN), via Fermi 40, I-00044*

## Abstract.

The paper is considering an opportunity for the CERN/GranSasso (CNGS) neutrino complex, concurrent time-wise with T2K and NOvA. It is a preliminary description of a ≈ 20 kt fiducial volume LAr-TPC following very closely the technology developed for the ICARUS-T600, which will be operational as CNGS2 early in 2008.

The present preliminary proposal, called MODULAr, is focused on the following three main activities, for which we seek an extended international collaboration:

(1) *the neutrino beam* from the CERN 400 GeV proton beam and an optimised horn focussing, eventually with an increased intensity in the framework of the LHC accelerator improvement programme.

(2) A *new experimental area* LNGS-B, of at least 50'000 $m^3$ at 10 km off-axis from the main Laboratory, eventually upgradable to larger sizes. As a comparison, the present LNGS laboratory has three halls for a total of 180'000 $m^3$. A location is under consideration at about 1.2 km equivalent water depth. The bubble chamber like imaging and the very fine calorimetry of the LAr-TPC detector will ensure the best background recognition not only from the off-axis neutrinos from the CNGS but also for proton decay and cosmic neutrinos.

(3) *A new LAr Imaging* detector, at least initially with about 20 kt fiducial mass. Such an increase in the volume over the current ICARUS T600 needs to be carefully considered. It is concluded that a single, huge volume of such a magnitude is uneconomical and inoperable for many reasons. A very large mass is best realised with a modular set of many identical, but independent units, each of about 5 kt, "cloning" the basic technology of the T600. Several of such modular units will be such as to reach at least 20 kt as initial sensitive volume. Further phases may foresee extensions of MODULAr to a mass required by the future physics goals.

Compared with large water Cherenkov (T2K) and fine grained scintillators (NOvA), the LAr-TPC offers a higher detection efficiency for a given mass and lower backgrounds, since virtually all channels may be unambiguously recognized. In addition to the search for $\theta_{13}$ oscillations and CP violation, it would be possible to collect a large number of accurately identified cosmic ray neutrino events and perform search for proton decay in the exotic channels.

The experiment might reasonably be operational in about 4/5 years, provided a new hall is excavated in the vicinity of the Gran Sasso Laboratory and adequate funding and participation are made available.

*(April 9,2007)*

[A] Corresponding author: Carlo.Rubbia@cern.ch

Table of contents.

# 1.— General considerations.

## *1.1. Physics introduction.*

The understanding of neutrino has recently advanced remarkably with the observation that they have masses and that oscillate between each other. Oscillations arise in analogy to the CKM matrix for hadrons since the neutrino species $\nu_e, \nu_\mu, \nu_\tau$ do not have specific masses, but are a combination of the mass eigenstates $\nu_1, \nu_2, \nu_3$. Two of these oscillations, namely $\theta_{12}$ related to $\nu_e \leftrightarrow (\nu_\mu, \nu_\tau)$ and $\theta_{23}$ related to $\nu_\mu \leftrightarrow \nu_\tau$ have been experimentally observed by SK[1]. A third oscillation type characterized by $\theta_{13}$, occurring around $\Delta m_{23}^2$ has not been observed. The observation of a non-zero value of $\theta_{13}$ will open the way to the ordering of the neutrino masses and a determination of the CP violation phase $\delta$ in neutrino oscillations. CP violation in the lepton sector will be necessary in order to understand why matter is dominating over anti-matter in the Universe. The very small but finite values of the neutrino masses require the existence of right-handed neutrino species and more generally neutrinos appear to be related to physics at an extremely high energy scale, beyond studies with accelerator beams.

It is also possible that in addition to the indicated three types of neutrino species, other species could exist, oscillated by the $\nu_e, \nu_\mu, \nu_\tau$. There is unconfirmed evidence for the existence of this type of “sterile” neutrinos from the LNSD experiment [1] at Los Alamos National Laboratory. A search for evidence for sterile neutrinos is being pursued by MiniBooNE [2] at FNAL and ICARUS-600T [3] at LNGS.

## *1.2. Comparing present and future detectors toward $\theta_{13}$.*

First generation long baseline neutrino experiments are currently operational at K2K [4] over a baseline of 295 km, at FNAL and at CNGS with baselines of about 730 km. These developments should be further exploited in Japan and presumably also in the USA and Europe with some second generation experiments of much higher sensitivity, to become operational around 2010-2015. This requires major improvements both in the beam and in the detector mass and performance.

The present detectors at FNAL (MINOS) [5] and CNGS2 (ICARUS) [3] are respectively a Iron-Scintillator sandwich of 2.5 cm iron and 4.1 cm wide scintillator strips with 5.4 kt total, 3.2 kt fiducial (MINOS, two modules) and a liquid Argon detector of a slightly lower mass of about 600 t of sensitive volume (ICARUS).

It is important to underline that in practice these two detectors have roughly comparable discovery potential in many channels because of the much higher resolution capabilities of LAr-TPC when compared with Fe-scintillation sandwich. The main beam requirement is the average target power of the incoming proton beam (POT) that are presently comparable for the CERN SPS[2] at 400 GeV and the FNAL main Injector at 120 GeV, with about 170 kWatt on target. It is foreseen that a major improvement programme at FNAL will increase the beam intensity to up to 1MWatt of beam power and even beyond.

---

[1] The experiment OPERA-CNGS1 is intended to observe explicitly the $\nu_\tau$ appearance (M. Guler et al., [OPERA Coll.], CERN/SPSC 2000-028, SPSC/P318, LNGS P25/2000).

An experimental proposal under consideration with a target date of circa 2012 is the NOvA [6] experiment, a totally active, fine grained scintillator of about 25 kt. In comparison with MINOS, NOvA will have (1) a much greater mass; (2) a better identification of the electron type neutrinos, with a sampling of 0.15 r.l., compared to 1.5 r.l. for MINOS; (3) about 80% of the mass is active, when compared to 5% for MINOS; (4) the beam is located off-axis, in order to increase the number of events which are most sensitive to $\theta_{13}$, namely $E_\nu = 2 \pm 1\ GeV$; (5) a much higher intensity neutrino beam, corresponding to 5 ÷ 25 x $10^{20}$ POT/y at 120 GeV, although this number is still subject to some uncertainties. The nominal quoted value is 6.5 x $10^{20}$ POT/y at 120 GeV.

We consider here the possibility of a substantial and equivalent upgrade of a LAr-TPC detector for CNGS, having in mind competition and timetable comparable to the ones of NOvA [6] and of T2K [7]. We keep in mind that the key process is the observation of the oscillation driven $\nu_\mu \to \nu_e$ events. As already pointed out, the use of the imaging capability of the LAr-TPC ensures a much higher discovery potential than in the case of scintillator (or water) detectors, i.e. a comparable sensitivity may be achieved with a much smaller sensitive mass.

The scientific community at large is presently considering also conceptual designs for huge, “ultimate” detectors in the order of one or more hundreds of kiloton [8] [10], with huge costs, comparable to the ones of the LHC experiments: some R&D efforts are presently going toward this long distance goal [9]. Amongst them one has described a bi-phase LAr-TPC detector of a mass of 100 kt with amplification in the gas phase (GLACIER) [10], a huge liquid scintillator of about 50 kt (LENA) [8] and a water Cherenkov counter of 440÷730 kt (MEMPHYS) [8]. In USA and Japan two analogous projects (UNO and HyperKamiokande) have been proposed [8].

But, no doubt the next practical steps for the period 2011÷2013 are still of more modest magnitude, of the order 20 ÷ 30 kt fiducial mass. Two programmes are already under development in Japan and in the US, namely the T2K combination of a new high intensity 50 GeV proton accelerator aiming at the well known SK water Cherenkov detector [7] and the NOvA scintillation detector of similar fiducial mass with neutrinos from the 120 GeV Full Energy Injector at FNAL [6]. They both are intended to operate with off-axis neutrino beams and for an optimum neutrino energy window of 2 ± 1 GeV.

The present paper is considering opportunities for the CNGS neutrino complex after the completion of the present OPERA/ICARUS phase, which should be completed by about 2011, concurrent time-wise with T2K and NOvA.

---

[2] However the fraction of time dedicated to neutrino beam is smaller at CERN than at FNAL The assumed efficiency for the parasitic operation at CERN is ≈ 50%, corresponding to a nominal 4.5 x$10^{19}$ POT/y, equivalent to 400/120*4.5 $10^{19}$ = 1. 5 x$10^{20}$ POT/y at FNAL and 120 GeV. The neutrino event rate is of 3000 ev/kt/y for FNAL and 2800 ev/kt/y for CERN.

## 2.— The next LNGS neutrino detector.

### *2.1. General considerations.*

The T600 detector is now readied in Hall B and it is expected to become operational before the end of 2007. We will collect in the subsequent years a large number of beam associated and of cosmic ray events that will perfect the technology and provide a rich amount of experimental physics. *The T600 is therefore considered a necessary step toward the realisation of any much larger LAr-TPC detector*. Running of the T600 will provide absolutely essential experience, which is required in order to develop sensibly such a “next step”. Evidently, a number of modifications are required in order to ensure the scalability of a detector to much larger sizes. These are presently under active consideration.

The present CNGS proposal is focused on the following three main activities, for which we seek a larger international collaboration:

1. *a new neutrino beam* configuration derived from the existing horn focussing and the existing proton beam line from the 400 GeV SPS, eventually with an increased intensity in the framework of the LHC related accelerator improvement programme. Relatively modest changes in the neutrino beam focussing of CERN will produce a nearly optimal beam configuration.
2. *A new experimental area*, eventually enlarged in future phases, which we indicate as LNGS-B to be realised about 10 km off-axis from the main laboratory, away from the protected area of the Gran Sasso National Park, without significant underground waters and with a minimal environmental impact. A provisional location is under consideration, corresponding to about 1.2 km of equivalent water depth. The high event rejection power of the LAr-TPC detector will ensure the absence of backgrounds not only for the neutrinos from the CNGS but also for proton decay and cosmic neutrinos.
3. *A new LAr-TPC Imaging underground* detector made of several modular units, each of about 5 kt fiducial mass. As a first step, a total of about 20 kt will be realised with appropriate safety requirements and along the lines of the vast R&D work carried out over the last decades by INFN and other International Institutes and culminating in the actual operation of the T600, foreseen in 2008 with the ICARUS experiment. This programme may eventually be improved further on with additional modules, depending on the developments of the programmes with and without accelerators.

The forthcoming operation of the T600 detector in the real experiment LNGS-B will represent the completion of a development of the LAr-TPC chamber over more than two decades. As it is described in this paper, the operation of the T600 evidences the large number of important milestones which have been already achieved in the last several years, opening the way to the development of this new line of modular elements and which may be extrapolated progressively to the largest conceivable LAr-TPC sensitive masses.

As described later on, the new detector will maintain the majority of components that have been developed with industry for the T600. The detector should be easily upgraded in the far future to a larger scale, depending on the potential physics goals.

The off-axis physics programme is not making obsolete the on-axis searches, presently concentrated on the T600, which both contribute to a wider physics programme and may also profit of the advances offered by the MODULAr concept.

*2.2. A modular approach of the LAr-TPC detector.*

Conceptual designs have been described in the literature [11] with a single LAr container of a huge size, up to 100 kt. But already in the case of containers of few thousand ton the geometrical dimensions of most types of events under study (beam-$\nu$, cosmic ray-$\nu$, proton decays) are relatively confined, i.e. much smaller than the fiducial volume. Hence increasing the container's size does not appreciably affect the acceptance in fiducial volume of each event and *introduces no significant physics arguments in its favour*. We believe that there are instead serious arguments for which such an huge size approach cannot be easily realised in practice and that suggest instead the use of a modular structure of several separate (identical) vessels, each one however of the size of a few thousands ton.

In case of an accidental leak of the ultra-pure LAr, the amount of liquid that is spoiled is proportional to the actual volume of the container. Segmentation is therefore useful in overcoming events due to poisoning of the liquid. In the case of a major damage of the detector, the liquid can be provisionally transferred to another container. An additional, reserve vessel of the order of 100 kt is, on the other hand, not realistic. In addition, the safety requirements of an underground vessel are strongly dependent on its size.

One of the most relevant features of LAr-TPC is its ability to detect accurately ionisation losses at the percent level. Over the very large volume, the inevitable in-homogeneities in electron lifetime due to even modest variations in purity of the LAr produce very large fluctuations in the actual value of the collected charge and hamper the possibility of charge determination along the tracks. Therefore we have chosen to use a modular approach of sufficient size in order to reduce the effects due to the non-uniformity of the electron collection due to the emergence of negative ions, which impose a reasonably short maximum drift distance of each gap.

As it will be discussed further on, it has been assumed that a reasonable sensitive volume should be of 8 x 8 $m^2$ cross section and a length of about 60 m, corresponding to 3840 $m^3$ of liquid or 5370 t of LAr. The drift length is 4 m. A field shaping grid should be added in the middle of the HV gap in order to reduce the effects due to the space charges to a negligible level. A reasonable three-plane wire pitch for such a large container should be of the order of 6 mm, twice the value of the T300.

*2.3. Double phase LAr-TPC signal collection?*

Several years ago [12] the ICARUS collaboration had studied a double phase Noble gas arrangement, in which ionization electrons from the tracks are drifted from the liquid to a superimposed gaseous phase. Electrons were further accelerated and ionised with the help of a grid, like in an ordinary gaseous TPC, before being collected by the readout wire planes. Dark matter searches in Argon (WARP) [13] and in Xenon (XENON) [14] have profited of this technique.

Essentially the same idea has been also envisaged some time ago for very large (≥ 100 kt) monolithic LAr-TPC detectors. In the specific case of GLACIER [10], it has been proposed a very large electron drift length of 20 m at 1 kV/cm in LAr, corresponding to a drift voltage of as much as 2 MV, about a factor ten larger than the one discussed in the present proposal. If one assumes that the free electron lifetime is at least 2 ms [10], one expects an attenuation of the free electrons due to ion recombination in the impurities, presumably Oxygen, of as much as a factor ~150 after 20 m. This residual free electron component cannot be directly recorded electronically (as in ICARUS) over the whole drift distance and must be therefore first amplified by the proportional gain profiting of the gaseous phase. In practice, taking into account the large capacitance of the extremely extended read-out electrodes (up to 70 m) and consequently of the larger noise signal from the input FET, gains of the order of $10^3$ are typical required values. The instrumentally increased dynamic range of the signals collected must take in full account this huge dynamic factor along the drift time extent.

Therefore in the double phase arrangement, all three different types of charged particles have to be simultaneously considered, namely (1) the initial free electrons from Argon which are attenuated over the distance by as much as two orders of magnitude, (2) the accumulated negative ions from recombination by impurities and (3) the positive Argon ions especially from the multiplication near the wire in the gas. The ion speeds in liquid are extremely slow, typically far less than 2 mm/s[3] at 1 kV/cm (with 2 MV over 20 m the drift time is ≈ 10’000 s, i.e. about 3 hours!).

It has been demonstrated in WARP and XENON experiments that free electrons can overcome the liquid to gas barrier in the presence of a sufficiently strong electric field of a few kV/cm. It is expected that both the positive ions travelling from gas to liquid and the negative ions travelling from liquid to gas, because of their larger masses and hence smaller speed, will be ultimately trapped and accumulate at each side of the liquid-gas boundaries for a so far unknown period of time.

The free electrons crossing the double ion layer cloud are presently under study with the 2.3 litre WARP detector underground at the LNGS. There is some preliminary evidence already in this small detector that space charges due to ion crossing at the boundary may introduce additional fluctuation in the electron ionisation signal. This will introduce a substantial worsening of one of the most relevant features of LAr-TPC, namely its ability to detect accurately ionisation losses.

The consequence for a detector of the size of 70 million litres and a diameter of 70 m, is an enormous extrapolation, which requires a very extensive R&D. The phenomenon of transfer of ions through the interface is expected to be rather complicated, not well understood and it has not been conclusively measured experimentally [15]. In the case relevant to GLACIER, it is likely that some trapping times may be ultimately occurring, but experimental studies are needed to assess how much. This is an absolutely crucial point prior to a successful, practical realization of a huge dual phase experiment in Argon.

---

[3] We remark that on such a timescale of hours convective motions inside the vessel become very relevant and they may seriously modify the time of drift of positive ions in either direction.

A first step is the 5 m long prototype ARGONTUBE [16] under study on surface in Bern, which will allow to experimentally verify these hypotheses and prove the feasibility of detectors with long drift paths, representing a very important milestone in the conceptual proof of the feasibility of the *dual phase detector* in Argon.

It is therefore concluded that at least at the present stage of the LAr-TPC, the *single phase geometry* which has been already very well developed experimentally [3] is vastly preferable and associated to a drift path length which could minimize the extent of the negative and positive ions. Negative ions in this configuration are smoothly drifting and directly captured by the collecting wires with a negligibly small signal (the electric signal is proportional to the drift speed, a factor $\approx 10^{-5}$ smaller for ions). Positive ions are straightforwardly collected at the cathode. For a sufficiently small drift volume, like the one described in the present proposal, the electric field distortions due to the slow ion motion can be made to be negligible.

*2.4. A simplified structure for the modular detectors.*

The structure of the detector has been considerably streamlined in order to reduce the number of components, its cost and increase the reliability of the system. The modular structure permits to repeat the initial engineering design of the prototype to a series of several subsequent units, reducing progressively their costs and their construction time.

Clearly the main aim of the detector is the one of filling and maintaining over many years a very large amount of ultra-pure LAr in stable conditions inside a dedicated underground cave, within very rigid safety conditions. The initial filling procedure is determined by the supply rate currently provided by the supplying industry. We believe that the maximum rate available in the European market is of the order of 200 $m^3$/d. At the rate of 100 $m^3$/d, the initial filling of the required about 25 kt of ultra-pure LAr (corresponding to a volume of 17'000 $m^3$) is about 170 days, which is acceptable. At the commercial cost of 0.7 Euro/l, the value of the LAr for the initial filling is about 12 MEuro, which is also quite acceptable. The mechanical structure of each of the modular units should be as simple as possible, keeping the costs of the various components commensurate to the relatively modest initial investment for the LAr.

The wire arrangement is scaled out from the industrial realisation of the existing ICARUS-T600, which is taken as a reference design. As well known ICARUS is made of two identical modules (T300). In each of the T300 made of two readout planes and a high voltage plane in a double gap configuration, the three readout planes have coordinates at 0° and ± 60° with respect to the horizontal direction. *This identical geometry is "cloned" into a larger modular detector, with the linear dimensions scaled by a factor 8/3 =2.66, namely the cross sectional area of the planes is now 8 x 8 $m^2$ rather than 3 x 3 $m^2$*. The wires in the longitudinal direction were originally 9.4 m long with the wire planes subdivided in two equal segments. In the next step the length will be quantised also into two individual wire sets, but 25 m long, corresponding again to the ratio 25/9.4 = 2.66. The longer wires have a higher capacitance and the signal/noise ratio is significantly decreased (wires, of the order of 10*pF/m*; cables, of the order 50p*F/m*). This factor is compensated widening the pitch to 6 mm, to be compared to the previous 3 mm, doubling the $dE/dx$ signals. Therefore we expect signal/noise ratios which are rather similar to the ones of the T600, namely of the order of 10/1. As it will be discussed later on, in collaboration with industry (CAEN), over the last several years the electronic chain from "wire to computer" has been considerably improved in performance and reduced in cost[4].

Each wire is now observing a time projected volume which is a factor 2.66 x (4/1.5) x 2 = 14.2 larger than in the case of the T600 (wire length x drift length x wire pitch). Therefore the average LAr mass observed by each TPC readout wire is about 200 kg/channel. A 20 kt sensitive volume will then require of the order of $10^5$ wires.

At this stage the configuration of the modules may not be considered as absolutely frozen and a number of possible configurations are possible, as shown in Figure 1, maintaining as a reference a readout plane dimension of 8 x 8 $m^2$.

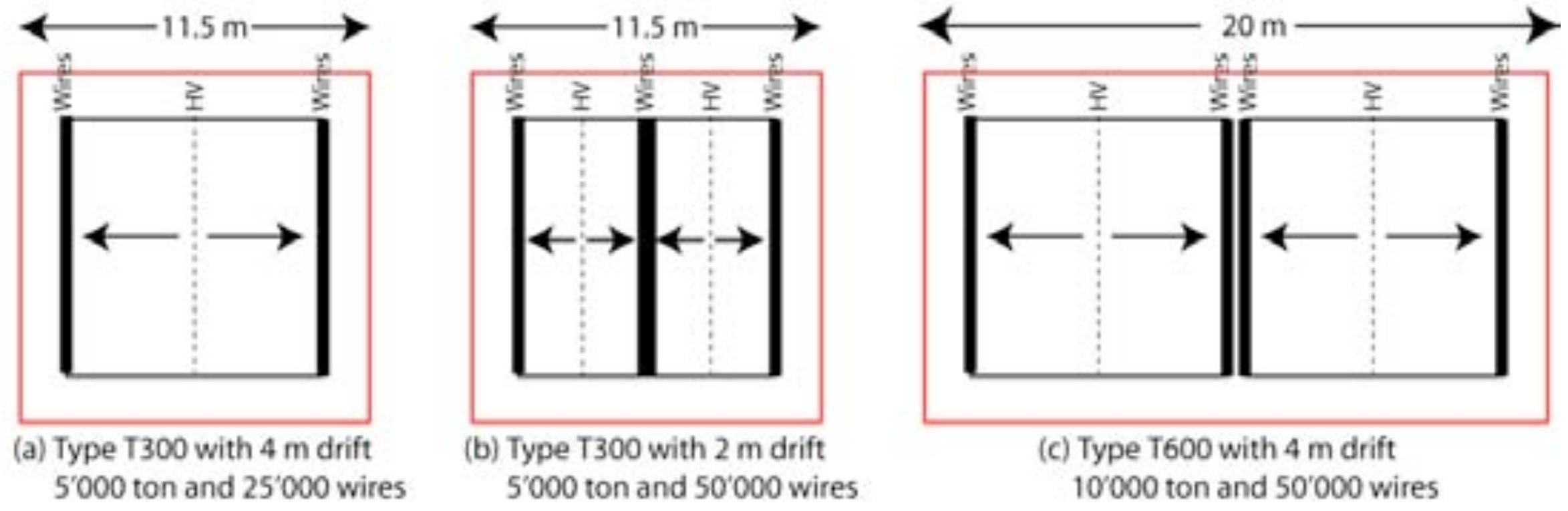

Figure 1. Various possible alternative arrangements for a modular unit. In (a) we show a scaled up T300 configuration with 4 m drift time and about 25'000 readout wires. The sensitive mass is about 5000 ton. In (b) the configuration is the same as in (a), except the drift distance has been reduced to 2 m and doubling the number of readout wires. In (c) we show a scaled up T600 configuration with a twin module, 4 m drift time and about 50'000 wires. The sensitive mass is then 10'000 tons.

Figure 1a represent the previously indicated basic configuration of a scaled T300 double gap arrangement. The nominal voltage of the T300 is 75 kV for the 1.5 m long drift, corresponding to a drift field of 500 V/cm, although the field-shaping electrodes have been currently operated without problems up to 150 kV. The engineering design for a T1200, never constructed, required a 3 m drift length. At the same nominal electron drift velocity (500 V/cm), for the present choice of 4 m drift, the HV would be 200 kV. However a significantly higher field, like for instance 350 kV, will shorten the drift time, ($\tau_{drift} \propto sqrt(E_{drift})$) and reduce correspondingly the requirements of purity for the LAr to the case already optimised of a T1200 with a 3 m drift.

---

[4] The estimated commercial cost of each channel "wire to computer" is now about 60 Euro. Assuming for each wire a sensitive volume of 200 kg, the levelized cost of the electronics is 60/200 = 0.3 Euro/kg, which is about ½ of the cost of the Argon procurement. The low cost achieved for such a sophisticated electronics is an additional argument in favour of a single phase LAr configuration with a relatively short drift gap distance, along the lines of the T600, rather than the double phase GLACIER arrangement.

In Figure 1b we have doubled the number of wire planes in order to reduce the drift distance to 2 m. In this configuration the halving of the drift time to the already successful configuration of the T300 is performed doubling the number of signal wires to 50’000, with a significant increase in the cost of the channels. In order to maintain an electron drift time exactly the same as the one of T300 since $\tau_{drift} \propto sqrt(E_{drift})$ we need an increase of the drift field to $75kV \times (2/1.5)^2 = 89kV$. Note that the HV of the T300 has been tested up to 150 kV without any problem. Although we believe that the drift distance can be safely extended to 4 m, this alternative shows that the choice of the electron drift length is not determinant. Solution 1b, eventually with an even higher drift field to reduce the maximum drift time to values below the ones of the present T600, is perfectly possible in case that some unforeseen problem may develop, obviously at the cost of doubling the number of electronic channels.

Finally in Figure 1c we show a scaled up T600 twin volume configuration, with 4 m drift time and about 50’000 wires. The two volumes are physically separated, but they are both kept in the same cryogenic volume. The total sensitive mass of one 1c module is then 10’000 tons. The new proposed halls of the LNGS and 20 kt could host two modules of type 1c in one container.

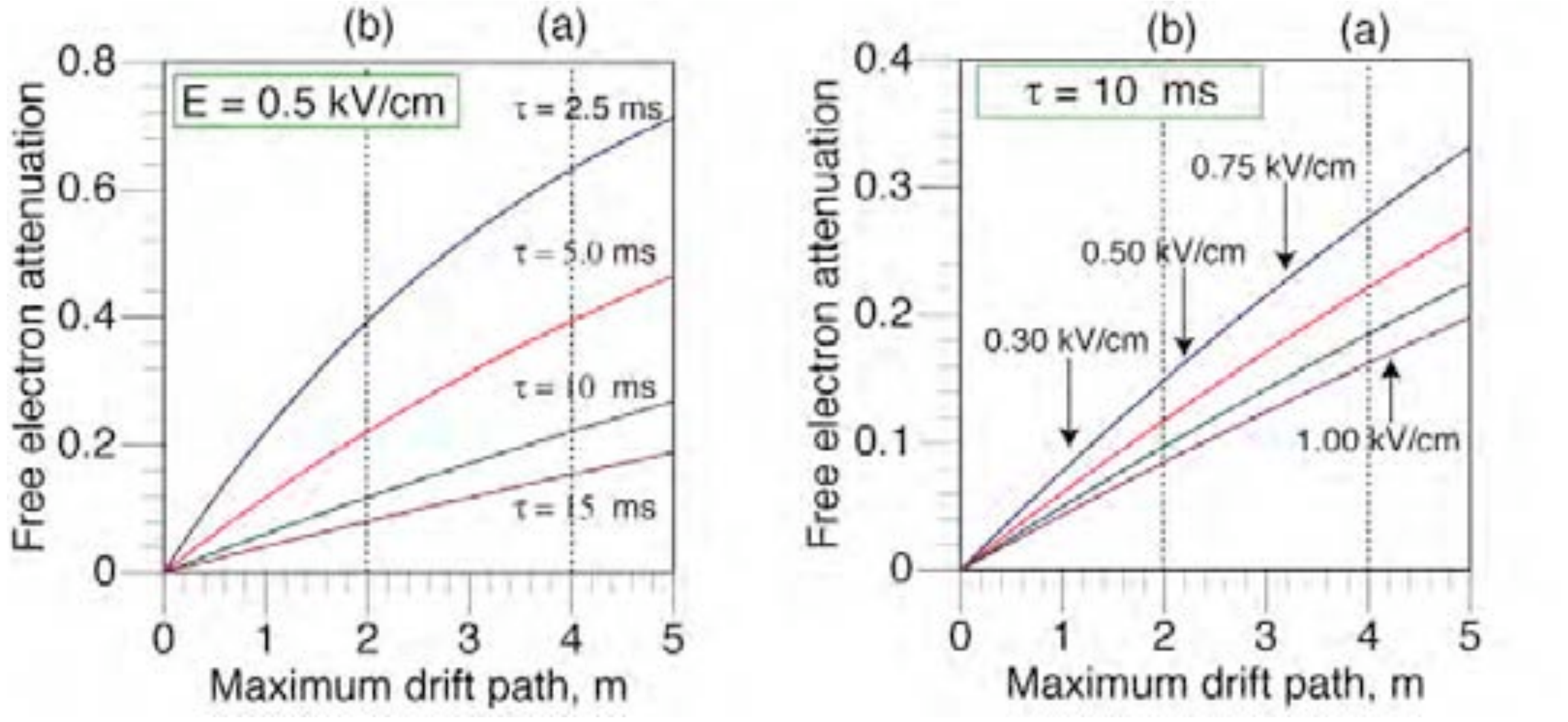

Figure 2. The maximum free electron attenuation into negative ions is shown for different values of the maximum electron drift path, respectively (1) for different values of the free electron lifetime at E = 0.5 kV/cm and (2) for different electric drift fields and a 10 ms electron lifetime. Cases (a) and (b) represent respectively the configurations of Figures 1a and 1b with 4 m and 2 m drift paths.

Considerable experience of the ICARUS collaboration has shown that free electron drift times $\tau_{drift}$ of several milliseconds are currently realised with commercial purification systems based on Oxysorb™. The effects on the electron attenuation are shown in Figure 2 where the drifting charge attenuation versus drift path at different electric field intensities are given for $\tau_{drift}$ = 10 ms and for different electron lifetimes at 0.5 kV/cm.

The longitudinal r.m.s. diffusion spread $\sigma_D$ after an electron drift path $x$ and moving at a speed $v_D$ is given by $\sigma_D = sqrt(2Dx/v_D)$, where $D = 4.06\ cm^2 s^{-1}$. In more practical units, $\sigma_D \approx 0.9 sqrt[\tau_{drift}(ms)]\ mm$. For a drift field of 0.5 kV/cm and a 4 m path the average value is $\langle \sigma_D \rangle = 1.1\ mm$ and the maximum value is $\langle \sigma_{D\max} \rangle = 1.6\ mm$.

The new mechanical structure, which has been highly streamlined, is essentially made of only three main mechanical components:

(1) An *external insulating vessel* made of two metallic concentric volumes, filled in between with perlite (see Appendix for details). Perlite is a mineral which is vastly used industrially, the world consumption being of the order of 2 million tons annually[5]. The environmental aspects of perlite are not severe: mining generally takes place in remote areas, and airborne dust is captured by bag-houses, and there is practically no runoff that contributes to water pollution. In order to ensure an adequate thermal insulation, about 1.5 m thickness is required, corresponding to over 3000 $m^3$ for a container. The bottom-supporting layer is made out of low conductivity light bricks. The specific heat loss is 3.86 $W/m^2$ for a nominal thermal conductivity of 0.029 (0.025-0.029) W/m/K. This is significantly smaller than the specific heat loss of the T600. Taking into account the dimensions of the vessel, the total heat loss is 8.28 kW. At present in the LNGS the cryogenic plant of ICARUS T600 is made of 10 units, each with 4 kW of (cold) power. Three of such units ($\leq$ 12 kW) should be adequate to ensure cooling of the walls of the vessel during normal operation. Evacuation of the perlite is therefore unnecessary.

(2) A linear *supporting holding structure* frame with wire planes at each lateral side and the high voltage plane at the centre. The photomultipliers for the light trigger are also mounted on this frame behind the wire planes. The structure of the planes is identical to the one already developed for the T600, except that only one wire out of two is installed in order to go from 3 mm to 6 mm pitch. The inner structure of the huge container is therefore extremely simple, being primarily a linear wire structure along the edges of the container, the rest remaining essentially free of structures.

(3) The *liquid Ar and N2 supply and refrigeration,* provided with cooling and purification both in the liquid and gas phases, with an appropriate re-circulating system to ensure that the whole liquid is moving orderly inside the vessel volume to unsure uniformity of the free electron lifetime.

*2.5. The new experimental area.*

The ICARUS-T600 detector is located inside the Hall B of the LNGS laboratory in an appropriate containment tank constructed above the floor of the Hall. The *new experimental*

---

[5] The present production in Greece, where vast resources are available, is about 500 thousand t/y. The estimated cost is about 40 $/ton.

*area*, which we indicate as LNGS-B, to be realised about 10 km off-axis from the main laboratory, away from the protected area of the Gran Sasso National Park, without significant underground waters and with a minimal environmental impact. A provisional location is under consideration, corresponding to about 1.2 km of equivalent water depth. The high event rejection power of the LAr-TPC detector will ensure the absence of backgrounds not only from the neutrinos from the CNGS but also for proton decay and cosmic neutrinos.

The total volume excavated for the original LNGS was of about 180’000 $m^3$. It is foreseen that the new LNGS-B could be about one half of this volume, namely initially about 50’000 $m^3$. However, as a difference from the main LNGS, the shapes of the cavities, rather than being vast, general purpose halls, are tailored to the specific experiment.

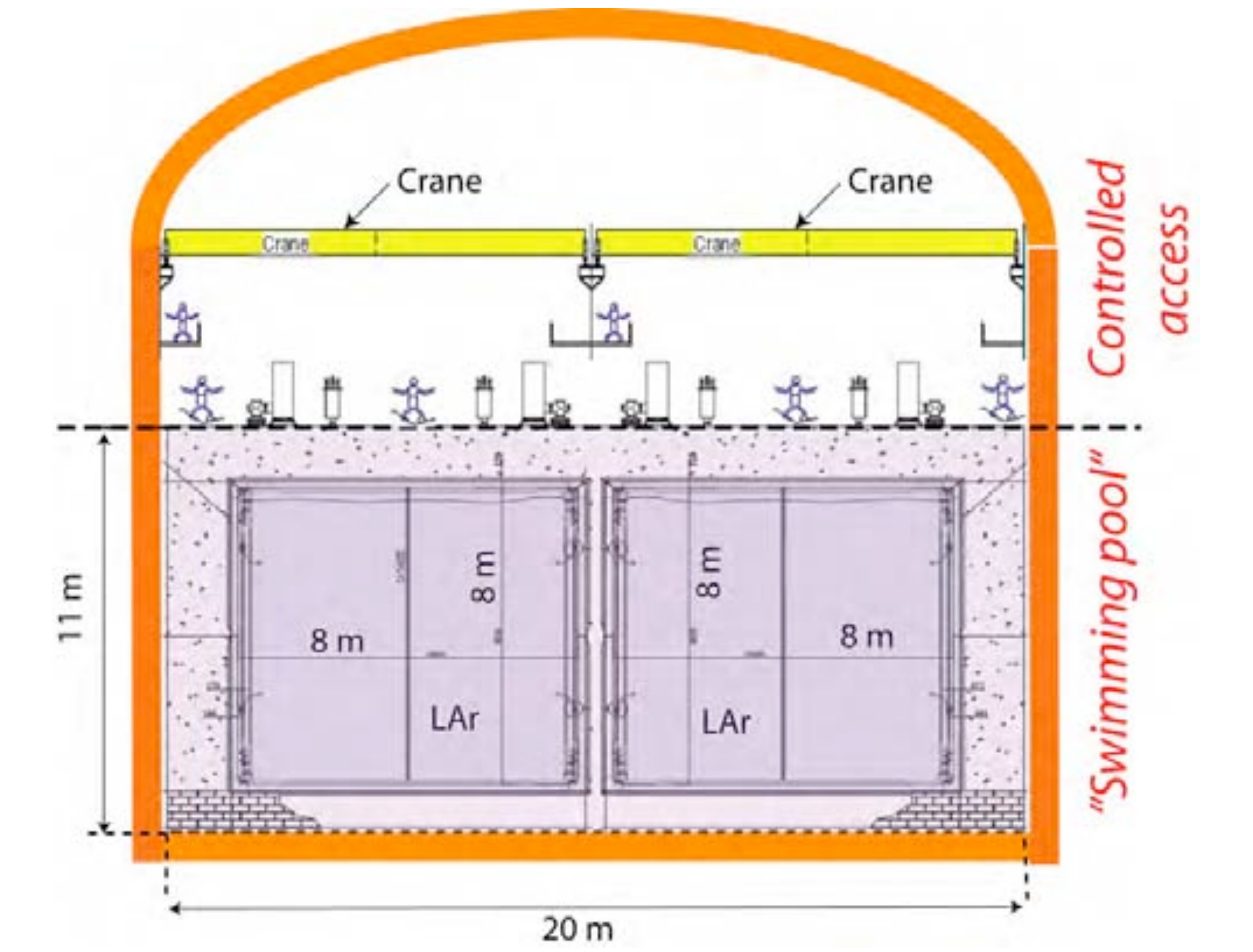

Figure 3. Indicative cross section of the T600 “clone” in the dedicated “swimming pool like” underground hall. The lower part is made of two twin separate LAr containers made of Aluminum extruded structures, thermally stabilized with forced N2 circulation. Outside the structure an about 1.5 m thick perlite wall provides spontaneous, passive heat insulation. The region on top of the “swimming pool” is accessible to auxiliary equipments. Personnel access is strictly controlled.

Each modular detector unit is located in an appropriate “swimming pool” cave in the rock concentric to the perlite walls, where the liquid tank is contained: therefore there is no realistic possibility of leak outside the walls of the rock for any foreseeable circumstance. The worst case is the total loss of external cryogenic cooling both of N2 and of Ar. Therefore the tank will spontaneously warm up in contact with the heat leaks of the surrounding components through the 1.5 m thick perlite wall. Assuming a heat leak rate of 10 kWatt, the LAr evaporation rate is of 220 kg/h, negligibly small with respect to the 5 kt of the stored LAr tank. Therefore the tank will remain stable in its liquid form for any specified length of time. More generally there is not even the most remote possibility to provide from the environment around the cavern a sufficiently large amount of heat in order to cause a catastrophic evaporation of a massive amount of LAr.

For a configuration of the type 1c, the cross section is 11 x 20 $m^2$ (shown in Figure 1) and the length is about 60 m. Different containers may have entirely separate halls since the event containment is anyway very good. An exhaust pipe is necessary in order to evacuate the evaporated liquid into the atmosphere in case of an accidental leak, although a risk analysis will certainly show that the probability of such events is very small.

*2.6. Initial filling procedures for the chamber.*

In the present T600 the vessel is evacuated in order to inject ultra pure Argon. The new detector, in view of its large size is very hard to evacuate and a new method has to be applied. The idea is to perform successive flushing in the gaseous phase in order to attenuate the presence of gases other than Argon with an approximately exponential chain. This method of flushing with pure Argon gas is widely used already in gaseous wire and drift chambers which are generally not evacuated. In the present case, additional problems may arise in view of the magnitude of the volume and the possibility of creating for instance “dead” spots, in which the gas may not circulate. A suitable small scale test dewar container is under construction in order to perfect the method.

In the idealised case of complete turbulent and continuous uniform mixing through the container, the transition air-argon is an exponential with a factor $\approx$ 1/2 at each passage. Therefore, in order to achieve an attenuation of the order of $10^{-6}$, 14 cycles are necessary. If instead the Argon is injected with little or no turbulence, for instance uniformly from the bottom with the extraction of initial air on the top, the transition argon-air moves orderly from the bottom to the top and only pure air exits from the top, producing a faster and more orderly transition. These are limiting cases and the efficiency of the actual filling will need some model studies and some hydrodynamic calculations to be perfected. Some preliminary considerations indicate that about 6 cycles may be necessary. The density of the gaseous Argon at room temperature is about 600 times smaller than the one of the liquid. Hence a ultimate gas purification of the order of $10^{-6}$ would correspond to an increment due to filling of the order of 2 x $10^{-9}$ (2 ppb) with pure liquid, which is adequate for the initial filling before local purification.

*2.7. LAr purification.*

In order to ensure a free electron lifetime adequate for the longest $\approx$ 3ms fly-path, a vigorous purification of the LAr must be kept at all times with filtering methods based on Oxysorb™ and molecular sieves. In analogy with what is currently performed with T600 and all previously constructed detectors, the purification is performed both in the liquid and in the gaseous phase. An improvement in the purification system is needed to enlarge in a significant way the TPC volume. New purification devices have to be implemented, possibly operating either near or directly inside the cryostat. They should be simple, robust and without moving parts, to guarantee a total reliability.

An order of magnitude of the liquid re-circulation rate needed to reach safely the running condition in few months (for example 60 days) could be of the order of several percent of volume per day. For a volume of 4000 $m^3$, this rate means a re-circulation rate of the order of 240 $m^3$/day (6%/day or purification cycle in 16.6 days), which is in the range possible with Oxysorb™[6].

In conclusion, some thousands cubic meters is a reasonable limit for a single TPC volume.  It could be cleaned, cooled and filled in few months and then kept completely operative (with an adequate LAr drift length) after few months. Altogether, such a detector could be put in operation in a reasonable period of time.

*2.8. Photo-multipliers for light collection.*

Like in the case of T600 a number of photomultipliers located behind the readout wires are used to provide a t=0 trigger. This is particularly important for the cosmic rays and proton decay events in which no starting signal is provided, but could be as well very useful in order to tag events coming from the CNGS beam. The technique already used in the case of the T600 consists in glass phototubes with a thin deposit of wave-shifter in order to record the scintillation light from the LAr. A significant contribution is also due to the Cherenkov light emitted directly in the visible by relativistic particles.

*2.9. Electronic readout and trigger.*

The ICARUS T600 detector has a DAQ system ($5{\cdot}10^4$ channels) designed at University of Padova/INFN, engineered and built by CAEN. It has proven to perform quite satisfactory in the test run performed in Pavia during summer 2001.

The electronics has been described in various papers and technical notes. We remind here that it is based on an analogue front-end followed by a multiplexed AD converter (10bit) and eventually by a digital VME module that performs local storage and data compression.

In the following, starting from the experience gained in the T600 operation, we discuss performance and limits of the actual system with the aim of improving its characteristics in view a multi-kton TPC with a number of channels in the order of several times $x\ 10^5$.

Since 1988, in the ICARUS proposal, the main characteristics of the signals were described and subsequently they were confirmed by tests on small chambers and eventually by the operation of the T600. In a multi-kton TPC we can foresee wires (or in general electrodes) with a pitch larger than the 3*mm* used in the T600. A reasonable assumption would be a 6*mm* pitch that will allow using most of the tooling already built and designed for the T600.

---

[6] The standard rate of a single Oxysorb™ pack is about 120 $m^3$/d. Therefore two of such units are sufficient for the chosen size of the vessel.

The capacitance associated to each channel will be determined by the capacitance of the wires, in the order of 10*pF/m*, in parallel with the capacitance of the cable, in the order 50*pF/m*. Let's assume in the following discussion ~600 *pF* as a reasonable number for 10*m* electrode wires and average 8*m* of cable.

The dominant noise in a high capacitance detector is the series noise $e_{sn}$ (voltage noise) linearly increasing with the input total capacitance ($C_D$) while the parallel noise (current noise) contribution is proportional to the shaping time of the signal.

We propose to use the present IC taking into account that due to the need of spares for the T600 a silicon run of the specific BiCMOS process of 6 wafers has been recently made. Each wafer, taking into account the known yield, contains some 12000 good circuits which means 24000 channels. The total number of channels that could be equipped is about $140*10^3$. These wafers, kept in inert atmosphere, can be easily packaged in very small cases (4x4 $mm^2$). A R&D program is also proposed for the development of a hybrid sub module hosting eight or more channels of amplification and eventually, as it will be described later, also the analogue to digital converter. At present a revision of the ICARUS analogue electronics is underway with the aim of further improving the front-end performance.

In the T600 collaboration a novel technique for the realization of feed-throughs has been developed. INFN holds a patent (RM2006A000406) for this technology that allows easy realization of feed-through with high density vias and different shapes.

The ADCs work at 20Mhz sampling rate, interleaved so the 10bit digital output has a 40Mhz frequency that means that each channel is sampled every 400ns. The power dissipated is significant: 500mW. The required bandwidth taking into account that two sets of 16 channels are merged in a 20 bit word, is 800Mbit/s for 32 channels.

The main feature of the new design is to move into digital domain all the conversion process at a very early stage and to exploit the use of numerical digital filtering techniques. The final quality of the converted data is highly dependent on the sampling frequency and numerical filtering.

The trigger system will divide the detector volume in sub-volumes to cope with the data acquisition rate required by shallow depth location. The basic structure will reproduce the one already implemented in the T600 and it will be based both on analogue signals from wires (sums of set of wires) and scintillation light detected by PMs inside the liquid Argon. All together will merge with the DAQ architecture taking into account that time resolution required is low (must be compared with drift time) and anyway the absolute time will be associated to each triggered event.

### *2.10. R&D developments.*

The increase of the active LAr volume of about one order of magnitude with respect to T600, the streamlining and simplification of the mechanical structures and the new developments previously described in the structure of the detector require some specific R&D developments, which are not of very substantial nature and could be implemented in parallel with the detailed engineering design of MODULAr. These developments (see Figure

4) are intended to simulate on a small scale the basic innovations with respect to the present T600, namely:

(1) The filling process starting from air to pure LAr, taking into account the motion of the gas, optimising the inlet and outlet geometries and minimising the number of cycles.

(2) The thermal convections of the LAr, in order to optimise the temperature gradients and to ensure a convincing circulation in all regions of the dewar, both in the cooldown phase and in the stationary state.

(3) The outgassing rate and the recirculation processes required in order to achieve the required electron lifetime

(4) The geometry of the compact re-circulators both in the liquid and in the gaseous phases.

The cryostat is based on a foam glass platform (the simplest solution, industrially well tested and implemented) and is surrounded by a perlite-insulated walls, about 1-meter thick.

This approach for the thermal insulation should be cheaper even if bulkier. The instrumentation must be well developed in order to be able to determine both thermal and convective measurements in a variety of conditions.

Finally, also electronics, starting from the positive experience of the T600 may require some specific development mainly on the layout improvement of the analogue part.

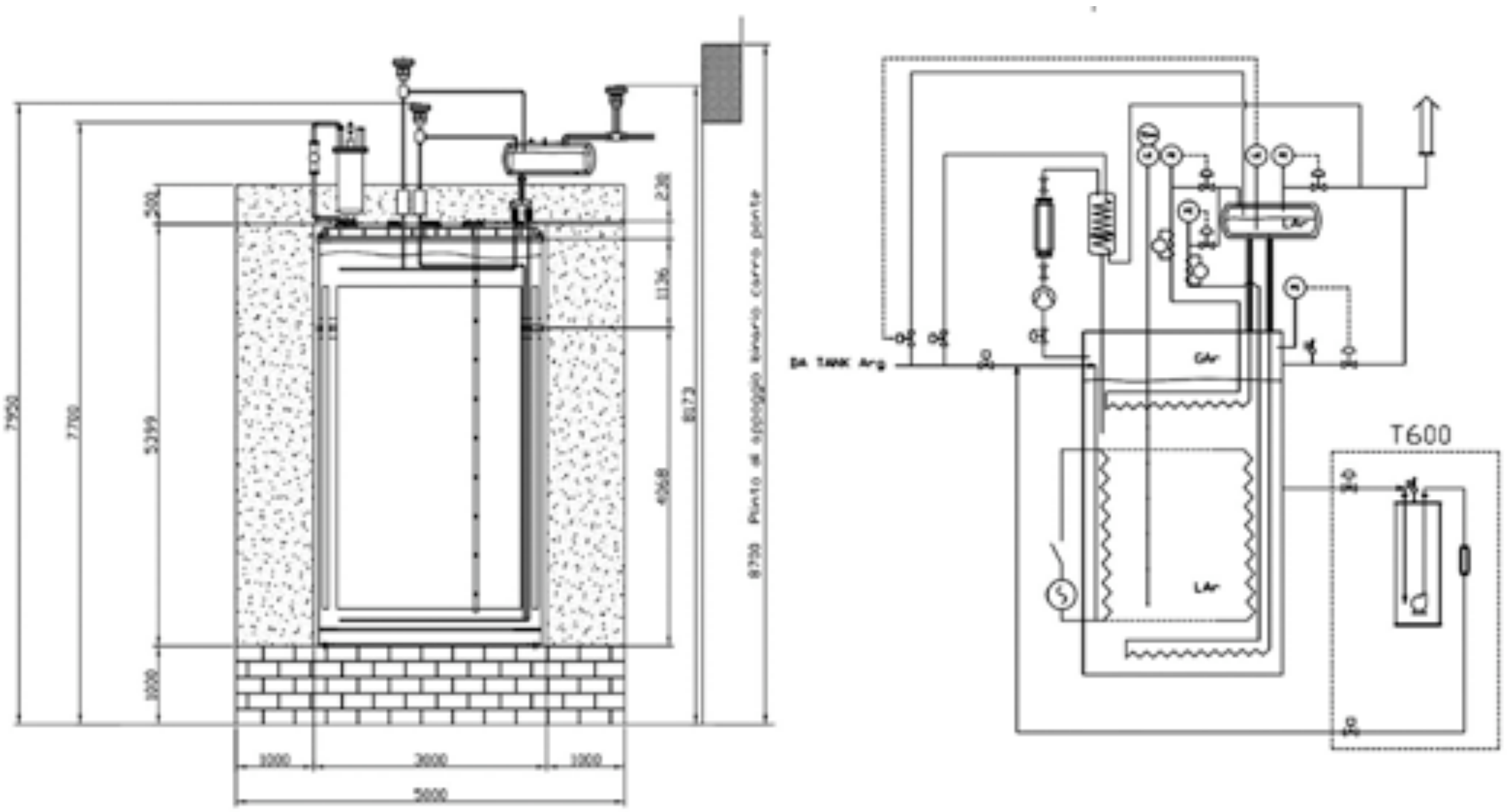

Figure 4. Layout of the test unit. This small unit is necessary in order to certify all the major changes with respect to the present T600 unit. The cross section is 5 x 5 $m^2$ and the height is about 8 m.

## 3.— The new low energy, off-axis neutrino beam (LNGS-B).

The primary goals of future experimental programmes in Japan (T2K), US (NOvA) and our present LOI at CNGS are related to the so far unknown angle $\sin^2(2\theta_{13})$, as a pre-requisite for a non zero CP-violation phase $\delta_{CP}$ in the lepton sector. They are all based on a neutrino beam with horn focussing and a fine grained detector of about 20-25 kt of fiducial volume.

As it is well known, the highest sensitivity occurs at distances corresponding to the maxima and minima of the cosmic neutrino oscillations, namely at the energy interval (2 ± 1) GeV for neutrino distances of 830 km (120 GeV) and 730 km (400 GeV) respectively at the NOvA and CNGS detectors since the oscillation maximum for the CNGS baseline, assuming $\Delta m^2_{23}$ =2.5 $10^{-3}$ $eV^2$, is at 1.5 GeV. The T2K detector will be driven by a 50 GeV proton beam and a smaller distance of 295 km and therefore should require correspondingly smaller energies.

In order to optimize a horn driven, conventional neutrino energy spectrum it is now universally agreed that the next configuration may be obtained using a high energy beam and looking at an off-axis direction in order to shift the Lorentz boost of the relativistic beam toward the required energy range, typically in the case of the LNGS and FNAL with displaced path distances of several kilometers from the main beam axis.

As well known this detector will also produce a large amount of additional non-accelerator physics results, and in particular of (1) proton decay especially in the SUSY channels and (2) of cosmic rays neutrino events, with considerable improvements with respect to SuperK. In order to be able to record simultaneously (pulsed) accelerator and (continuous) non-accelerator events, the detector must be triggered by the inner photo-multiplier array, as already indicated by the T600.

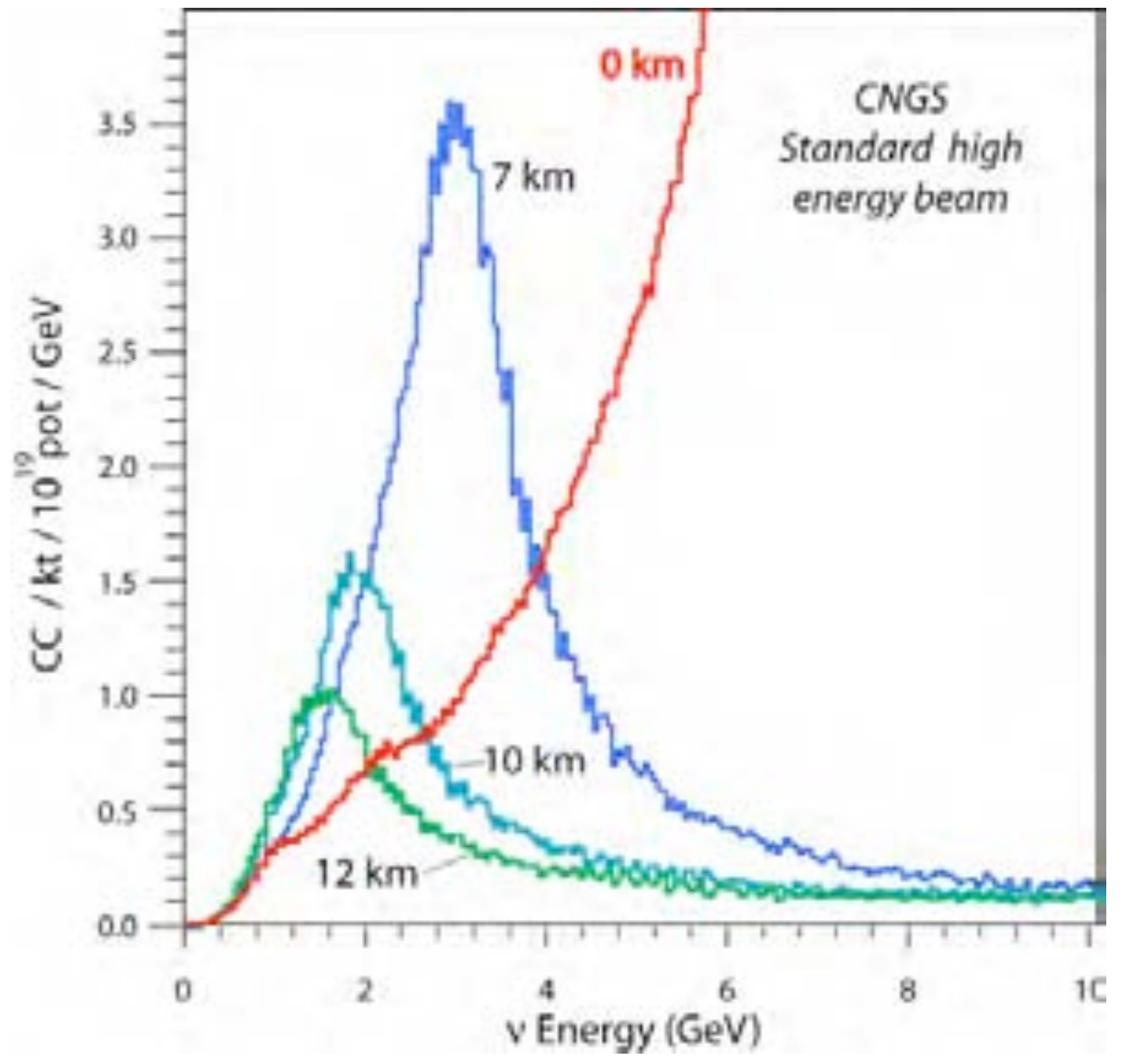

Figure 5. The present 400 GeV CNGS neutrino CC interaction signal in all its oscillated channels at 731 km, at three different distances off-axis and on-axis. For instance, the value for 12 km off axis corresponds to 16 mrad or 0.94°.

On a longer timescale it may be possible to realize the entirely new technology of beta-beams [17], where the purity of $\nu_e$ is extremely high, namely a $\nu_\mu$ contamination well below $10^{-4}$. Evidently the residual $\nu_\mu$ contamination due to neutral currents must also be reduced correspondingly. As shown in Ref. [18] the presently developed LAr technology will be capable of realising such required powerful identification capabilities. Therefore the massive LAr-TPC detector presently under consideration may continue to be used also in this ‘ultimate” phase, eventually increasing the number of modular units.

*3.1. The present high energy CNGS beam configuration.*

The present 400 GeV CNGS beam, optimized for τ appearance experiment, is a high-energy wide band beam with an average on-axis energy of 18 GeV. The horn/reflector optics is designed to focus in the forward direction secondary particle in the 20-50 GeV energy range with an angular acceptance of 20-30 mrad. The thin target is optimized to minimize pion/kaon re-interactions. The integrated $\nu_e$contamination is 0.6%. From comparisons with the previous WANF beam and the early observations with CNGS, the systematic uncertainty on the calculated $\nu_e$ /$\nu_\mu$ ratio is expected to be very good, of the order of 5% [19]. Therefore the intrinsic $\nu_e$ contamination, although significant in the search for $\nu_\mu \rightarrow \nu_e$ oscillations, may be very precisely calibrated.

In order to visualize the situation we consider the present CNGS beam geometry (Figure 5). Displacing the detector off-axis with respect to the neutrino beam direction by several kilometres at the location of the detector strongly reduces the shape of the neutrino energy spectrum, which, as expected, becomes progressively softer as the distance from axis is increased. The strong $\nu$ CC signal at 0 km is progressively shifted to smaller energies with a much smaller over-all rate, although significantly enhanced at the optimum energy.

*3.2. A new, low energy focussing layout.*

The standard 400 GeV CNGS optics as such is not optimized for $\nu_\mu \rightarrow \nu_e$ oscillation searches, neither on-axis nor off-axis. In particular, the small angular acceptance of the magnetic lenses, around 20-30 mrad, limits the neutrino fluxes at low energies. Optimization of the target and beam optics with low energy focussing has been calculated and it is in progress. It indicates that a design close to that proposed by the NOvA collaboration is also applicable in the case of CNGS. The main changes with respect to the present CNGS design are:

(1) A more compact (without air gaps) and thicker target, to increase tertiary production at low energy;
(2) Larger acceptance (up to 100 mrad) of the horns system in the 7 – 20 GeV pion energy (this momentum range offers a kinematically efficient production of 2 – 3 Gev neutrinos at 10 –15 mrad off-axis).

A shorter tunnel length (between one half and two third) could also be considered because it would marginally reduce the $\nu_e$ flux due to muon decays that primarily happen in the downstream part of the decay tunnel; this major upgrade of the civil engineering has not been considered in the present study.

In Figure 6 we show the expected event rate, for a hypothetical $\nu_\mu \rightarrow \nu_e$ oscillation equal to the present CHOOZ upper limit $\theta_{13} = 11°$, $\delta_{CP} = 0$ and the intrinsic $\nu_e$ contamination from the beam. Of course this is the highest expected signal and most likely the oscillatory signal is of much smaller size. Calculations are the result of a full beam simulation based on the FLUKA MonteCarlo code [20].

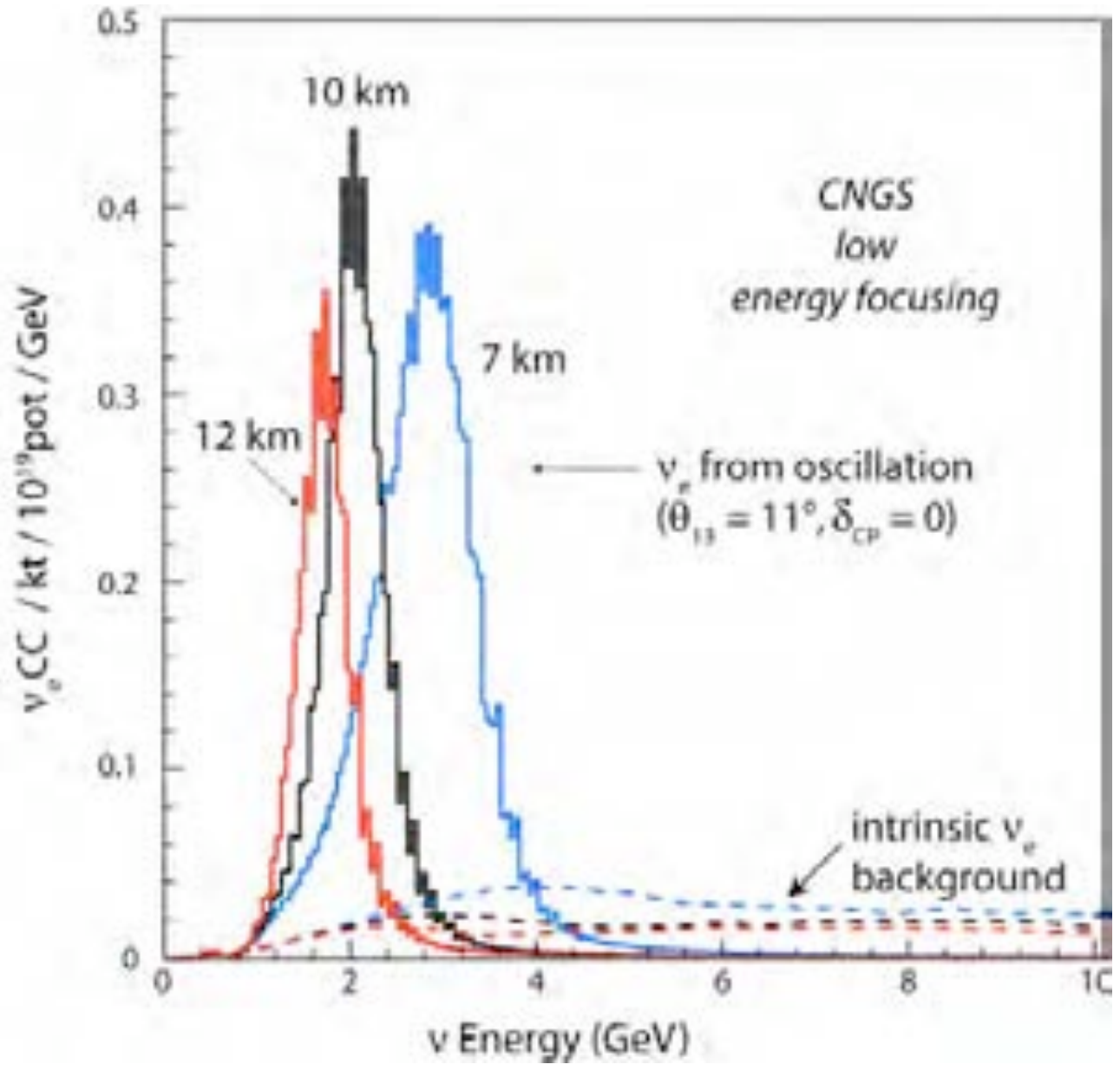

Figure 6. Calculated beam spectra for low energy focusing. Rates are estimated for $10^{19}$ protons and 400 GeV on target. Both the event rate for an hypothetical $\nu_\mu \rightarrow \nu_e$ oscillations equal to the present CHOOZ upper limit $\theta_{13} = 11°$, $\delta_{CP} = 0$ and the intrinsic $\nu_e$ contamination from the beam are shown.

Such new low energy CNGS optics (see Figure 7) is very similar to the one proposed for NOvA at 120 GeV, taking into account that the energies of the protons on target (POT) are in the ratio 400/120 = 3.3. Preliminary calculations at 14.8 mrad off-axis and at a baseline of 732 km demonstrate that the neutrino spectra obtained with a proton beam of 400 GeV are very similar in shape to those obtained with 120 GeV protons, except for the yield, substantially higher at 400 GeV.

The relative behaviour can be easily factorized in a good approximation. While the resulting neutrino beam intensity grows roughly proportionally with proton energy and *therefore the neutrino flux in the interesting beam energy region is roughly proportional to the beam power on target,* the detailed shapes of both the initial $\nu_\mu$ and of the $\nu_e$ intrinsic beam contamination are relatively unaffected by the proton energy and depend primarily on the choices of the target/horn system, the focused momentum range and the secondary particle acceptances.

The beam intensities at 400 GeV and 120 GeV are shown in Figure 7. The plot shows that in the focused energy range the ratio of the muon fluxes due to the higher proton energy is about 2.6, namely about 80% of the linear increase expected on the energy alone, slowly growing to the full proton energy factor for the higher energy tail.

An entirely similar effect is observed for the ratio of the $\nu_e$ intrinsic beam contaminations and therefore no appreciable difference is also observed in the $\nu_e$ background.

*3.3. Comparing the CNGS and NOvA neutrino beams.*

The CNGS neutrino beam is presently given for a rate of 4.5 x $10^{19}$ POT/y at 400 GeV. This rate is believed as generally insufficient for the future off-axis neutrino programmes. A vigorous programme is on its way at FNAL in order to accumulate in line with the NOvA's nominal assumption[7] a yearly rate of 6.5 x $10^{20}$ POT/y at 120 GeV. On the basis of the previous considerations (factor 2.6) this corresponds to approximately 2.5 x $10^{20}$ POT/y at 400 GeV, about a factor 5.5 larger than the present CNGS yearly performance. Evidently such a factor must be recovered at CERN.

[7] Ref [6]. chap. 11,page 75

We should compare more in detail dedicated repetition rates and energies at FNAL and CNGS. The nominal NOvA's is given for a beam intensity of 6.0 x $10^{13}$ ppp and a repetition rate of 1.5 s, corresponding to a cycle integrated beam power of 768 kW. The nominal values for CNGS today are two proton batches 50 ms apart, in total 4.8 x $10^{13}$ ppp, every 6 s, corresponding to a beam power of 512 kW, which is 2/3 of the FNAL value. Therefore the instantaneous cycles for fully dedicated accelerators are quite comparable.

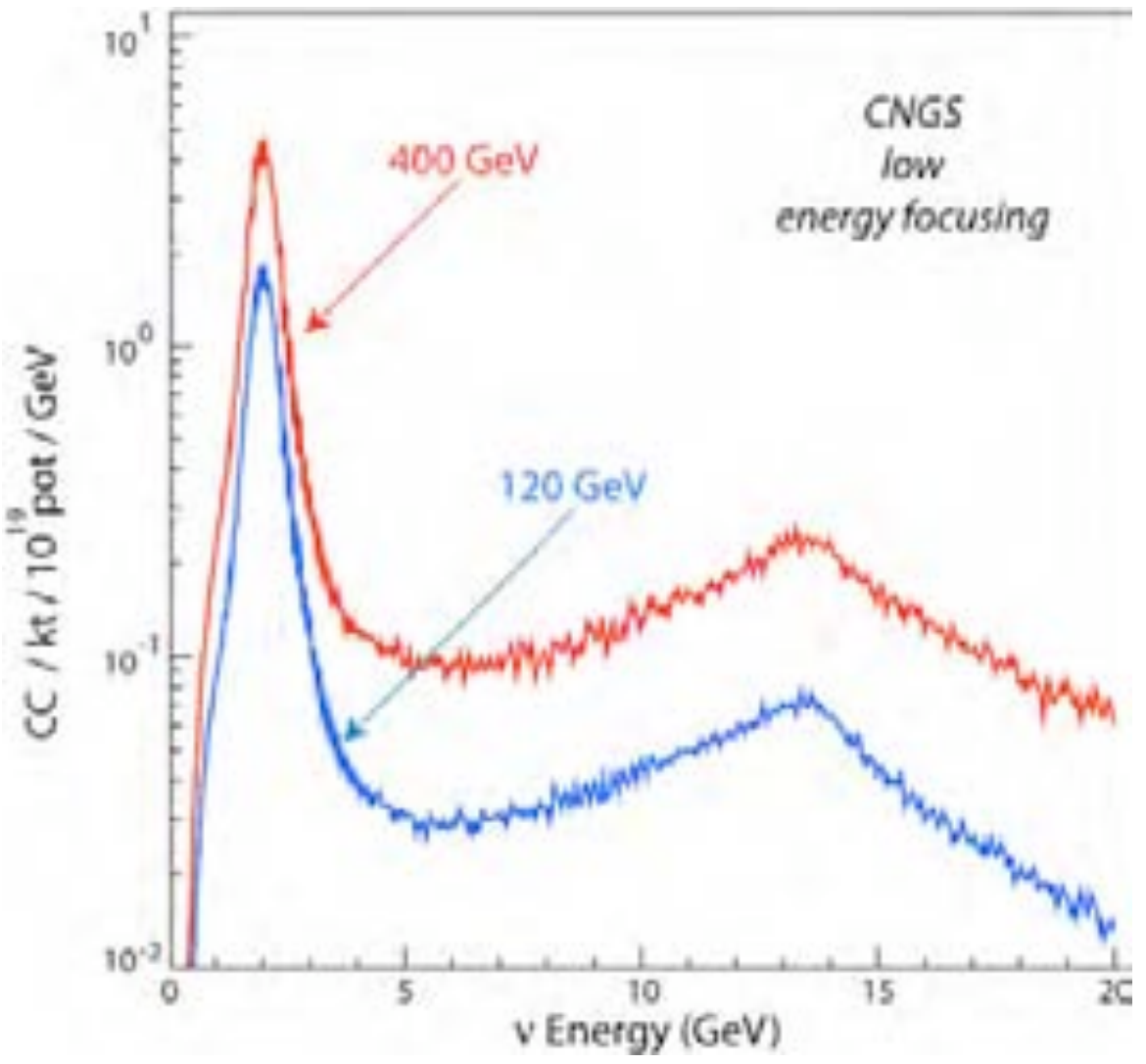

Figure 7. Muon neutrino CC event rates for $10^{19}$ protons at 120 GeV (FNAL) and 400 GeV (CNGS), calculated with the low energy focusing optics, 14.8 mrad off-axis and a baseline of 732 km and without oscillation mixing.

The differences (a factor 5.5 rather than 1.5) therefore come primarily because of the present CNGS parasitic operation with fixed SPS target, an efficiency factor of 55% in the operation of the SPS complex and a less intensive operation of the accelerator complex because of the LHC/fixed target sharing.

Such "human factors" should be supported in order to reduce the gap between the two accelerators. We can assume that in several years from now, a dedicated 6 s cycle rate for the neutrino beam and an efficiency factor of 80% rather than 50% may become possible. These factors should bring the integrated intensity to *1.2 x $10^{20}$ POT/y at 400 GeV*, corresponding to about 4.0 x $10^{20}$ POT/y at 120 GeV. This value is not far from the figure assumed by NOvA's of *6.5 x $10^{20}$ POT/y at 120 GeV*. With such improvements, presumably possible in several years from now, the present CNGS beam may become roughly competitive with FNAL, the neutrino fluxes at the detector being then in the ratio 1.6 to 1.

Table 1. Rates for 5 years, 20 kt and 1.2 $10^{20}$ POT /year. Oscillation with $\sin^2(2\theta_{13})=0.1$. The upper integration limit, $E_{lim}$, has been chosen to get the best sensitivity, $S/sqrt[bkg]$

| | $0 < E < E_{lim}$ | | | | | $0 < E < 10$ GeV | | |
|---|---|---|---|---|---|---|---|---|
| | $E_{lim}$ (GeV) | μ | e $[bkg]$ | Signal $S$ | S/√(bg) | μ | e $[bkg]$ | Signal, $S$ |
| *Present high energy configuration* | | | | | | | | |
| 7 km | 3.5 | 6200 | 34 | 190 | 33 | 1000 | 200 | 240 |
| 10 km | 2.5 | 2300 | 15 | 101 | 26 | 4600 | 160 | 130 |
| *Low energy focussing* | | | | | | | | |
| 7 km | 3.5 | 13000 | 70 | 390 | 47 | 17500 | 340 | 430 |
| 10 km | 2.5 | 5700 | 28 | 250 | 47 | 7800 | 230 | 280 |

Integrated CC event rates for 5 years at 1.2 $10^{20}$ POT/year in a 20 kt fiducial volume of a LAr-TPC detector are summarized in Table 1. Signal events have been calculated for $\sin^2(2\theta_{13})$=0.1 to allow easy scaling. The actual size of the $\nu_\mu \rightarrow \nu_e$ oscillation driven $\sin^2(2\theta_{13})$ is obviously unknown. In order to estimate the sensitivity for a small signal $S$ in the presence of a significant beam associated $\nu_e$ background $[bkg]$ we consider the quantity $S/sqrt[bkg]$ as a figure of merit as a function of the off-axis distance (see Figure 8).

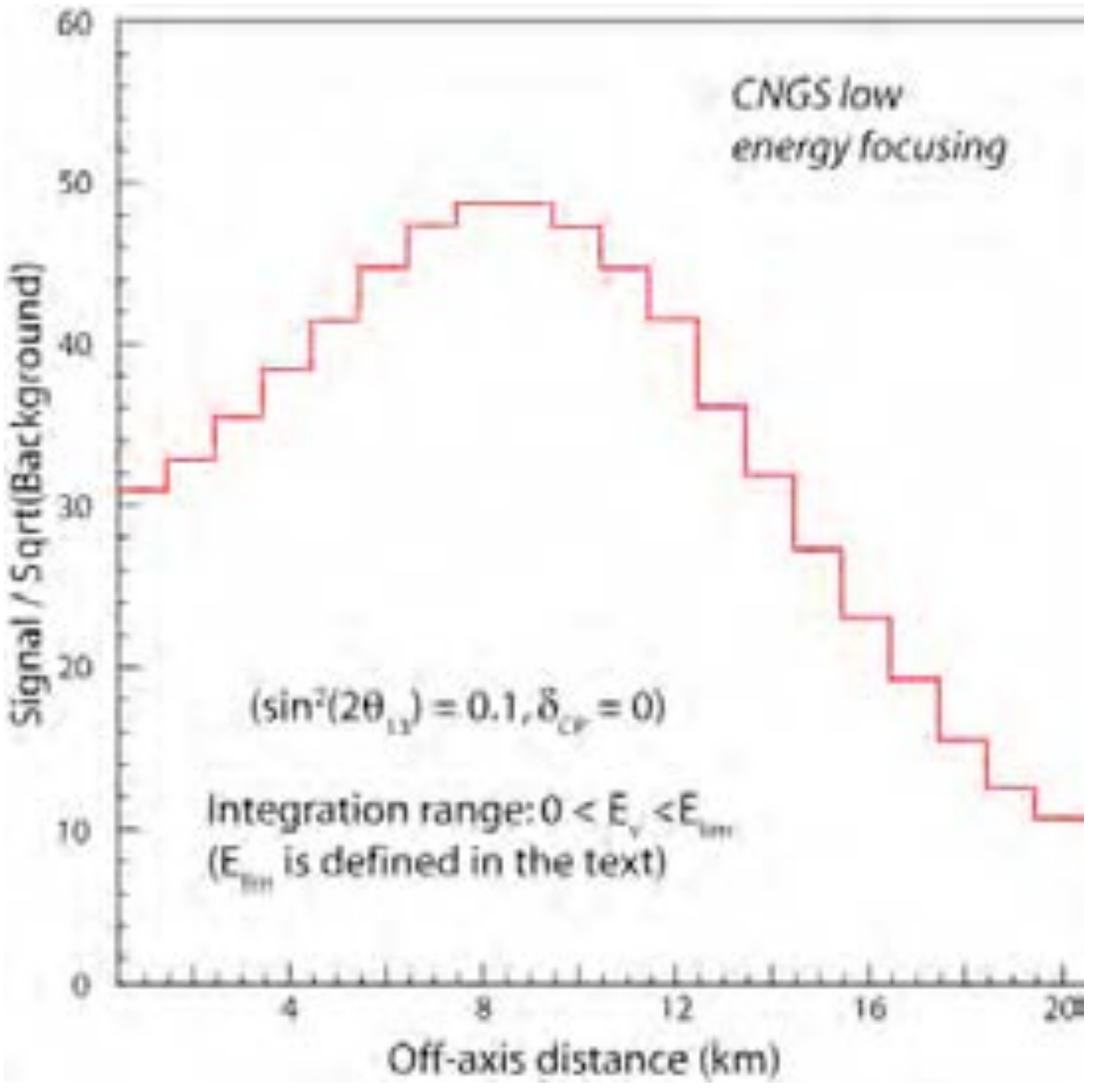

Figure 8. Sensitivity to oscillation expressed in terms of the $S/sqrt[bkg]$ (intrinsic plus tau), as a function of the off-axis distance. The plot indicates the off-axis distances 7 ÷ 12 km (0.55$^0$ ÷ 0.94°, 9.6 ÷ 16.4 mrad) are the optimum.

The upper integration limit, $E_{lim}$, has been chosen to get the best sensitivity, $S/sqrt[bkg]$. It varies with the off-axis distance because of the different spectrum shape of signal and background. As a reference the rates, integrated up to 10 GeV, are also shown for an estimation of the signal selection efficiency in the case of the optimum cut (80% in the $\tau$ optics vs. 90% in the low focus case).

Both at CERN and at FNAL increases in the proton beam intensity are conceivable with relatively modest efforts, but require additional money. A further increase to as much as a factor 4 has been considered at FNAL with a new 8 GeV Proton Driver, provided it will be built in the near future. Several increases of the accelerated intensity may be considered also in the case of CERN, especially based on the improvements of the CPS and in connection with the several LHC improvement programmes. In both cases the main limit may not be however the proton accelerator but rather the capability of the target/horns complex to withstand the required power.

### *3.4. Detection efficiency for $\nu_e$ CC events and NC background rejection.*

In the previous section, the NC background has been assumed to be negligible. This assumption is supported by extensive studies performed by ICARUS collaboration [3,21] both for beam and atmospheric neutrinos. In particular, the analysis of NC background rejection in a low energy on-axis neutrino beam [22] which has been extensively performed for the ICARUS configuration can be applied as well to the off-axis beam described in the present work. We summarize here this analysis and its results. A full simulation of the events in LAr-TPC was performed to study the background of neutral pions in both neutral current and charged current interactions, which may simulate a background of $\nu_e$ induced CC.

For each neutrino event was recorded the visible energy, defined as the energy deposited by ionization in the sensitive LAr-TPC volume, with the exception of the ionization due to heavily quenched recoils. The optimization of the fiducial volume was based on three criteria: energy reconstruction, electron identification and $\pi^o$ identification.

(1) *The energy reconstruction* due to non-containment of the neutrino events affects the signal/background ratio, especially when the signal is restricted in a narrow energy range. However, this non-containment is severe only for interactions occurring near the end of the detector and in a few centimetres lateral skin. A minimal cut of 50 cm in the longitudinal direction and 5 cm on the sides of the sensitive volume has been found sufficient to preserve the signal/background ratio. In the large modules as described in the present proposal, these cuts would reduce the fiducial volume by a (negligible) 2%.

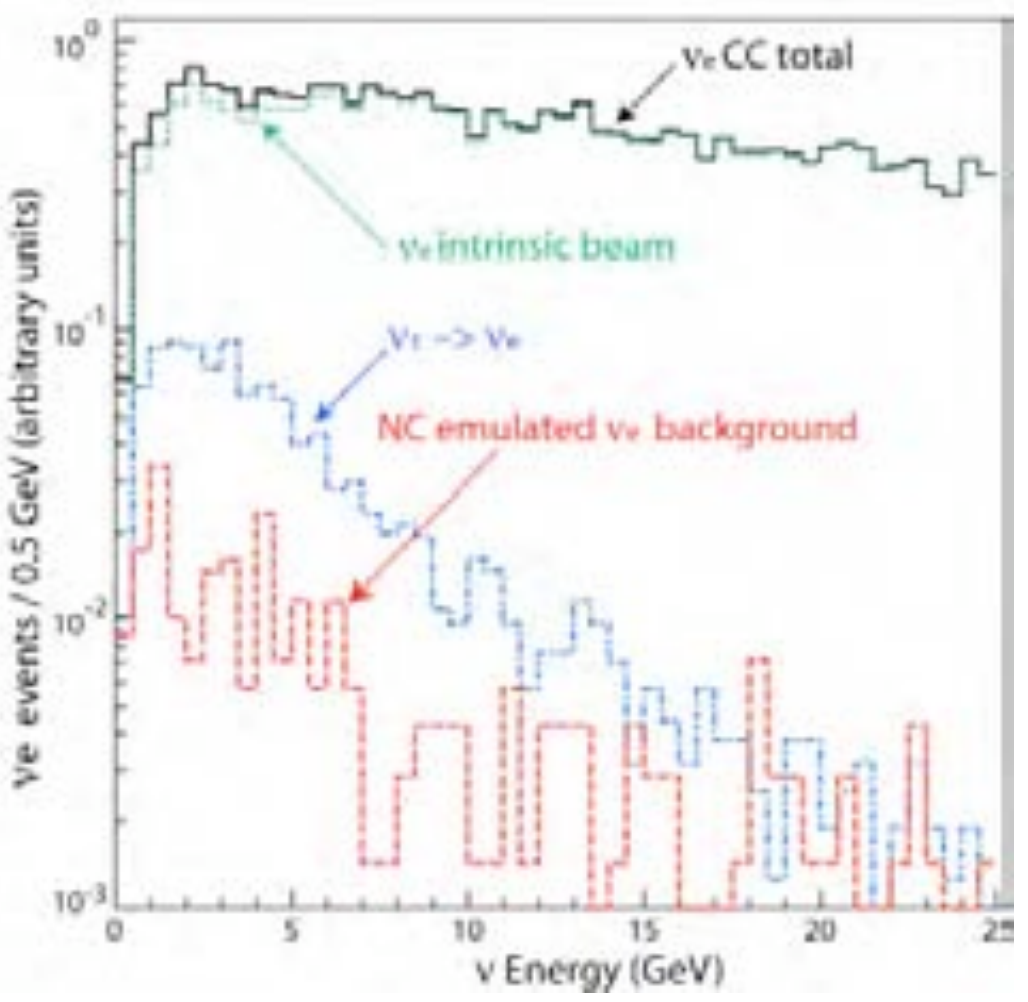

Figure 9. Background sources at LNGS on-axis with a low energy CNGS beam as studied for the ICARUS detector. This calculation is very similar to the expectations from the off-axis low energy beam, with the exception of the $\nu_\tau$ background, which is much smaller. The residual NC background in the LAR-TPC, after rejection of neutral pions, is about two orders of magnitude smaller than the intrinsic beam associated $\nu_e$ background.

(2) *Electron identification* is also assured under these geometrical cuts. Indeed, due to the directionality of the neutrino beam the probability that an electron escapes from the instrumented volume before initiating a shower is extremely small: only 2 % of the electrons "travel " through a LAr-TPC thickness smaller than 3 $X_0$, and 0.3 % travel less than 1 $X_0$ in the instrumented volume. Therefore we can safely assume to identify electrons with almost 100 % efficiency.

(3) $\pi^o$ *identification.* Neutral pions from $\nu_\mu$ NC events could be misidentified as electrons. But because of the superior imaging capability of LAr-TPC technology, all events where both photon conversion points can be distinguished from the ν interaction vertex can be rejected. To be conservative, in the ICARUS analysis only photons converting at more than 2 cm from the ν vertex were rejected. The remaining neutral pion background was further reduced by assuming that events where the parent $\pi^o$ mass can be reconstructed within 10 % accuracy are discarded. The effect of the last requirement obviously depends on the assumed fiducial volume. Only 4 % of $\pi^o$'s survive the cuts when all interaction vertexes are accepted, and a further decrease to 3% is obtained when the fiducial volume is restricted as described above. On the remaining photon sample we apply the results obtained in Ref. [23] on the possibility to discriminate electrons from photons on the basis of dE/dx. This method provides a 90% electron identification efficiency with photon misidentification probability of 3% at relatively low energies. The misidentification

probability is expected to decrease with energy due to the decreasing contribution of Compton scattering. A track length of 2.5 cm is sufficient to achieve the discrimination. After all cuts, the final $\pi^o$ mis-interpretation probability is 0.1 %, while the corresponding electron identification efficiency is 90 %.

The residual background, after all the above cuts are applied, is shown in Figure 9 in the case of the study performed for the ICARUS detector exposed at a CNGS beam on-axis [22]. Given the very similar energy spectrum and background contaminations, the same signal efficiency and NC rejection power can be applied in the case of the off-axis beams. The residual NC background, after rejection of neutral pions in the LAR-TPC is about two orders of magnitude smaller than the intrinsic beam associated $\nu_e$ background.

*3.5. Comparisons with NOvA.*

In the present letter of intent we consider the possibility of a substantial and equivalent upgrade of a LAr-TPC detector for CNGS, having in mind competition and timetable comparable to the ones of NOvA. We keep in mind that the key process is the observation of the oscillation driven $\nu_\mu \to \nu_e$events. As already pointed out, the use of the imaging capability of the LAr-TPC ensures a much higher discovery potential than it is the case of scintillator (or water) detectors, i.e. a comparable sensitivity may be achieved with a much smaller sensitive mass. The higher performance of LAr-TPC introduces important advantages with respect to NOvA, namely:

- the NOvA detector is mostly limited to elastic events while LAr-TPC may collect also all kinds of inelastic events. An elaborate MC simulation of the NOvA proposal[8] of the signals and backgrounds for oscillations using relevant parts of the MINOS experiment software, the NEUGEN3 neutrino interaction generator and the GEANT3 detector simulation show that the efficiency for accepting an event from $\nu_\mu \to \nu_e$oscillations is approximately 24%. This introduces about a factor 4 in rate with respect to LAr-TPC in which virtually all event configurations are identified, for the same fiducial mass;
- NOvA in contrast with LAr-TPC is may be contaminated by neutral current events that fake electron events, while in reality they are due to $\pi^o$. The background is typically about two-thirds from beam's produced from muon and kaon decay and one-third from neutral-current events. This increases the background of $\nu_\mu \to \nu_e$ events and increases further the level of the over-all discovery potential by a factor $sqrt(1.5/1) = 1.22$. We conclude that a ≈ 5 kt LAr-TPC detector should have performances comparable to the ones of NOvA. We underline that the NC cross sections are today only poorly known and therefore the magnitude of the effect, absent for LAr-TPC has to be carefully measured in separate experiments, at least as long the $\sin^2(2\theta_{13})$ signal is close to the sensitivity limit.

---

[8] Ref [6]. chap. 12, page 78

Similar considerations apply also about the water Cherenkov counter at T2K, where apparently more stringent cuts are necessary with a corresponding reduction of the rates in order to improve the sensitivity for the small signal $\nu_\mu \to \nu_e$ oscillation driven $\sin^2(2\theta_{13})$, so far obviously unknown, with respect with the $\pi^o$ related backgrounds.

*3.6. Evaluation of the beam associated $\nu_e$ background.*

The future CNGS low energy neutrino beam most likely will lack of an appropriate near detector. As a consequence the estimations of $\nu_e/\nu_\mu$ background ratio will be primarily based on the Monte Carlo simulation of the beam. The on-axis beam flux could however be determined experimentally (Figure 9) with the help of ICARUS-T600 neutrino detector presently in Hall B, measuring the neutrino spectra components with relatively high statistics (~100 $\nu_\mu$ CC events/kt/$10^{19}$POT, peaked at 7 ± 2 GeV). Once the agreement is confirmed on-axis, it may be possible to normalize and tune the simulation for the beam off-axis. The possibility of operating simultaneously the on-axis and the off-axis events, both with very similar LAr-TPC detectors, represent a very important correlation amongst the two measurements.

Experience with the CERN-WANF beam has demonstrated that FLUKA based calculations, compared with measurements of the $\nu_\mu$ spectrum, are able to provide an accuracy on the $\nu_e$ background normalization at the level of 3 % as well as a error on the spectrum shape better than 4% [24]. Similar considerations should bring precious new information for the future low energy off-axis neutrino beam.

When applied to the off-axis beam the overall errors are reduced since the kinematics of the meson decay confine the neutrino spectra in a much narrower energy range, hence the error associated with the spectrum shape knowledge is strongly suppressed.

*3.7. Comparing the ultimate sensitivities to $\delta_{CP}$ and $\sin^2(2\theta_{13})$.*

The value of $\delta_{CP},\sin^2(2\theta_{13})$ plane to be searched upon is currently unknown, although the experimental upper limit is 0.14. In order to be detectable, the number of $\nu_\mu \to \nu_e$ must substantially exceed the intrinsic $\nu_e$ beam contamination. For instance, even in absence of NC contamination, which is the case of LAr-TPC, equal rates of oscillations and of $\nu_e$ contamination (the results for FNAL and CNGS are quite similar) correspond to $\delta,\sin^2(2\theta_{13})$ ≈ 1.6 x $10^{-2}$.

As already pointed out, sensitivity to smaller values of $\delta_{CP},\sin^2(2\theta_{13})$ implies an accurate knowledge of the actual $\nu_e$ contamination and of the neutrino cross sections. The sensitivity to a non-zero value of $\delta_{CP},\sin^2(2\theta_{13})$ is then proportional only to the square root of the number of events.

The sensitivities for $(\delta_{CP},\sin^2(2\theta_{13}))$ plane at 3$\sigma$ for T2K, NOvA and the proposed future scenarios at CNGS, all based on 5 years of operation, 20 kt LAr-TPC detector 10 km off-axis are shown in Figure 10 with 1.2 $10^{20}$ POT/year at 400 GeV (1) and with a delivered proton intensity of 4.3 $10^{20}$ POT/year (2).

Calculations have been performed with the GLoBES code [25], assuming for the energy resolution a conservative value of 15% and a global 5% uncertainty on the beam composition.

As expected, the sensitivity with the new CNGS beam is definitely better than the one of T2K and NOvA.

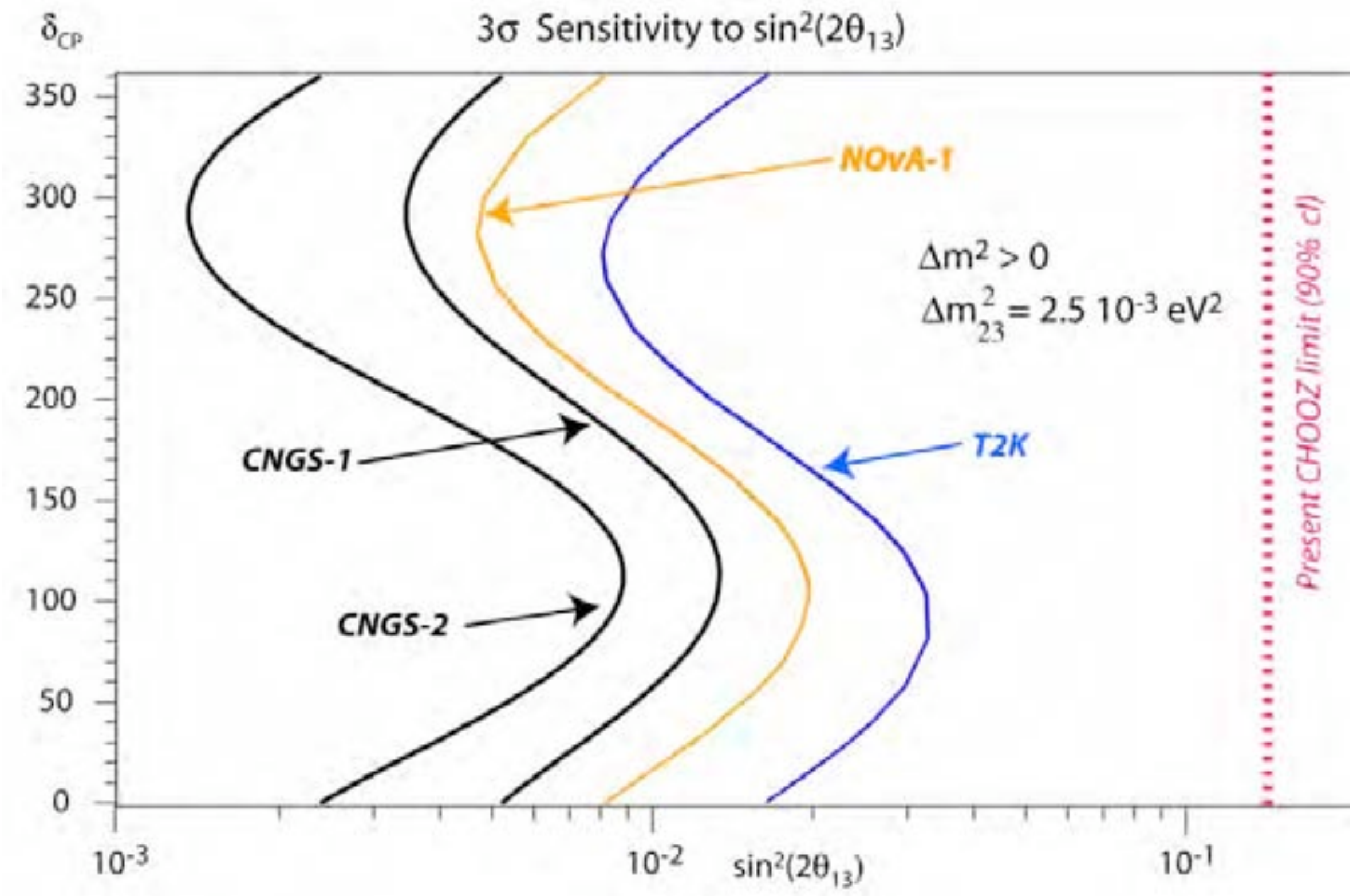

Figure 10. Comparison of the sensitivities for $\delta_{CP}$, $\sin^2(2\theta_{13})$ plane at 3σ, all based on 5 years of neutrino operation for T2K, NOvA-1 and two proposed future scenarios at CNGS, 20 kt fiducial LAr detector. The CNGS-1 beam configuration is for (1) 1.2 $10^{20}$ POT/year at 400 GeV, (2) a cycle integrated beam power of 512 kWatt, (3) a new target/optics configuration optimized for a low energy neutrino beam, (4) 10 km off-axis detector and (5) without substantial increases of the SPS performance. The CNGS-2 configuration assumes an hypothetical improvement of the SPS and CPS to 4.33 $10^{20}$ POT/year corresponding to 1.6 MWatt beam power. The intensity of the NOvA-1 experiment is 6.5 x $10^{20}$ POT/year at 120 GeV and a a cycle integrated beam power of 768 kWatt. The corresponding intensity improvements for NOvA-2 are not completely identified and therefore the sensitivity is not shown. . All cases are computed for Δm > 0. Note the much higher sensitivity offered by the LAr approach which ensures higher discovery potentials, since every type of event is then clearly recognized and identified, contributing to the determination of the oscillation phenomenon. As a comparison, for instance in the case of NOvA, only 24% of the events may be used and a sizeable contaminant of mis-interpreted neutral current events add to the intrinsic $\nu_e$ emission from the beam. Similar inefficiency considerations apply to the water Cherenkov counter at T2K.

## 4.— Tentative layout of LNGS-B.

As described, the off-axis arrangement implies the realisation of a separate underground cave at about a significant distance with respect to the main CNGS beam line. A preliminary study has been conducted[9] in order to identify the most appropriate new location, which we indicate with LNGS-B, keeping in mind a number of conditions:

- The underground laboratory may be at a depth, which is much shallower than the one of the main Laboratory. The high event rejection power of the LAr-TPC detector will ensure the absence of backgrounds not only from the neutrinos from the CNGS but also for proton decay and cosmic neutrinos. A depth of about 400 m of rock, corresponding to about 1.2 km of equivalent water depth has been chosen.
- The location of the experimental halls should be between 7 km and 12 km from the axis of the beam.
- The new laboratory should be out of the protected area of the Gran Sasso Park.
- The neighbouring rock should not imply any presence of significant underground water and a minimal environmental impact.

The general layout of the landscape across Gran Sasso massif side is shown in the top of Figure 11. Two potential locations at 10 km from the CNGS beam axis have been identified; location A is on the Teramo side of the mountain (close to L'Aquilano village), location B is on the L'Aquila side (close to Camarda village). Both locations fullfil the requirements of being outside the Gran Sasso National Park Area in a water free rock at a sufficient rock depth (400 m), providing an adequate shielding to cosmic radiation (as shown at the bottom-left of Figure 11); moreover they are easily reachable through the ordinary roads. Site A (shown at the bottom-right of Figure 11) is preferred because it requires a shorter access tunnel for a given depth.

The new cavern must have a gas exhaust of an adequate cross section to the surface and be organised with the instrumental entry from the top, well above the perlite insulated LAr-TPC container. The entrance to the cave may be made with a strong door which can be closed and undergo pressurisation in case of a major accident. In this case the only exit of the gas will be through the exhaust pipeline.

---

[9] The study of the choice of the new hall locations as well as the preliminary layout design have been commissioned by the Prof. E. Coccia (director of LNGS) and performed by Ing. Roberto Guercio (Univ. Roma I, Italy).

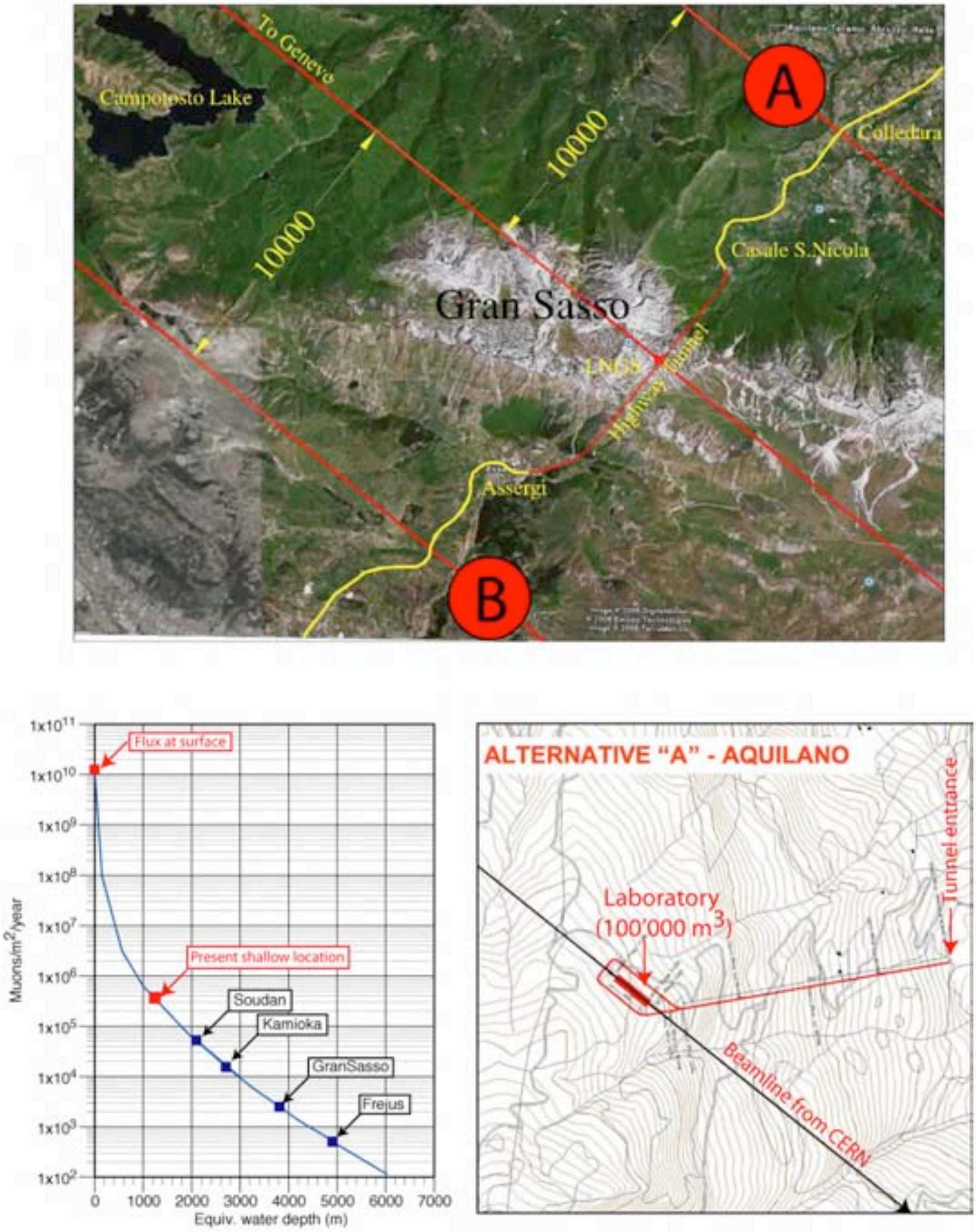

Figure 11. General layout for a potential new site to detect low energy off-axis neutrinos from the CNGS beam. In the top picture two potential locations, one at each side of the existing laboratory have been indicated with (A) and (B) located at 10 km off axis from the main CNGS beam, respectively in location Aquilano and Camarda. The natural muon background is shown next (equiv. water depth ≈ 1.2 km) and compared both with the flux at the surface and with several existing underground laboratories. The Aquilano location is also shown, since it offers the shortest distance of the tunnel. They are both outside the natural park and they can be easily reached from main roads.

## 5.— Conclusions.

The forthcoming operation of the T600 detector in the real experiment CNGS2 will represent the completion of a development of the LAr-TPC chamber over more than two decades and it opens realistically the way to truly massive detectors for accelerator and non accelerator driven phenomena. As it has been described in this paper, the operation of the T600 evidences that a number of important milestones have been already achieved in the last several years, opening the way to the development of a new line of modular elements and which can be extrapolated progressively to the largest conceivable LAr-TPC sensitive masses.

The new detector will maintain the majority of components we have already developed, in particular:

- The readout electronics and the data acquisition of the T600, which has been developed in collaboration with the CAEN company. The 50'000 channels of electronics already at hand are adequate for about 10'000 tons of sensitive mass of new modular elements.
- The signal feed-throughs for the very large number of signal wires, which have been developed for the T600 can be applicable directly to the new modular elements. The technology, which has been patented by INFN/Padova, has shown itself extremely reliable and capable of withstanding the extreme leakage rate of the many tens of thousands of feed-throughs at LAr temperatures.
- The original technology for the realisation of the wire planes and associated structures to withstand the very large changes of temperature during filling of the detector and which has been industrially realized by the company CINEL. A detailed engineering design, which has been developed for the module T1200 but never realized in practice, is also completely compatible and ready for the present design.
- An original and extremely robust technology for the readout wires has been developed, capable of withstanding the wide variations in temperature (-200 K) in the cooling and warm up phases. This technology has been developed in collaboration with industry. So far in the T600 not a single wire has broken, in spite of the many operations and of the transport on road for about 600 km. Although the wires are about 2.66 times longer, which does not constitute a problem, the same method will be cloned to the new modular elements. The realisation is simple, fast and cheap and it is realised with the help of an automatic machine.
- The high voltage feed-through and the appropriate > 100 dB noise filtering to remove the electric noise from the very small signals of the wire chambers has been developed and fully operated with no problem at 150 kV, which is twice the design voltage of the T600. The design voltage of the new modular elements is 200 kV.

- The purity of the LAr is generally well below the required level of 3 x $10^{-10}$ of equivalent Oxygen purity, which has required a dedicated technological development over the many years. The LAr is continuously purified both in the gas and in the liquid phase and circulated with adequate low temperature pumps. The purification system can be expanded in a straightforward way to become adequate to the new modular elements. A great deal of experience has taught us how to remove materials that are producing significant leakages.
- Photomultipliers, which are wave shifting the 128 nm Argon light into the visible, have been designed with the help of the EMI Company in order to ensure an appropriate photocathode efficiency at the LAr temperature.
- An appropriate method of high precision purity monitors, in order to monitor in real time the purity of the LAr.

As pointed out already the main domain of remaining developments, to ensure the correct realization of the new modular elements, is related to the streamlining and simplification of the mechanical structures, to the reduction of the overall costs and to the new developments, previously described, of the structure of the detector, which are:

- The use of perlite for the cryogenic structure. Perlite is vastly in use in the cryogenic industry and should represent no problem.
- The filling process starting from air to pure LAr, taking into account the motion of the gas, optimising the inlet and outlet geometries and minimising the number of cycles.
- The thermal convections of the LAr, in order to optimise the temperature gradients and to insure a convincing circulation in all regions of the dewar, both in the cooldown phase and in the stationary state.
- The outgassing rate and the recirculation processes required in order to achieve the required electron lifetime.
- The geometry of the compact re-circulators both in the liquid and in the gaseous phases.

The realization of these developments should not rise any insurmountable problem and the detailed engineering design of the first new modular elements should proceed smoothly and rapidly.

The experiment might reasonably be operational in about 5 years, provided a new hall is excavated in the vicinity of the Gran Sasso Laboratory and appropriate funding is made available.

## 6.— References

## 7.— Appendix. General comments on the use of Perlite.

Perlite is a generic term for naturally occurring siliceous volcanic rock. The distinguishing feature which sets perlite apart from other volcanic glasses is that when heated to a suitable point in its softening range, it expands from four to twenty times its original volume.

This expansion process is due to the presence of two to six percent combined water in the crude perlite rock. When quickly heated to above 870 C the crude rock pops in a manner similar to popcorn as the combined water vaporizes and creates countless tiny bubbles in the softened glassy particles. It is these tiny glass-sealed bubbles which account for the amazing lightweight and other exceptional physical properties of expanded perlite.

The expansion process also creates one of perlite's most distinguishing characteristics: its white color. While the crude perlite rock may range from transparent to light gray to glossy black, the color of expanded perlite ranges from snowy white to grayish white.

Expanded perlite can be manufactured to weigh from 32 kg/m$^3$ to 240 kg/m$^3$. Because of its unique properties, perlite insulation has found wide acceptance in the insulating of cryogenic and low temperature storage tanks, in shipping containers, cold boxes, test chambers, and in food processing.

Perlite insulation suitable for non evacuated cryogenic or low temperature use exhibits low thermal conductivity throughout a range of densities, however, the normal recommended density range is 48 to 72 kg/m$^3$, that is about 1/20 of the density of water. In addition to its excellent thermal properties, perlite insulation is relatively low in cost, easy to handle and install, and does not shrink, swell, warp, or slump. Perlite is non-combustible, meets fire regulations, and can lower insurance rates. Because it is an inorganic material, it is rot and vermin proof. As a result of its closed cell structure, the material does not retain moisture.

Thermal conductivity varies with temperature, density, pressure, and conductivity of the gas which fills the insulation spaces at mean temperature -126 C°, but it is typically in the interval 0.025-0.029 W/m/K.

In the present design the volume of perlite is 3928 m$^3$, corresponding to a mass of 235 t at an expanded density of 60 (48 to 72) kg/m$^3$. At the thickness of 1.5 m and a temperature difference of 200 K, the specific heat loss is 3.86 W/m$^2$ for a nominal thermal conductivity of 0.029 (0.025-0.029) W/m/K. This is significantly smaller than the specific heat loss of the T600. Taking into account the dimensions of the vessel, the total heat loss is 8.28 kW. At present in the CNGS the cryogenic plant is made of 10 units, each with 4 kW of (cold) power. Three of such units ($\leq$ 12 kW) should be adequate to ensure cooling of the walls of the vessel during normal operation.

A higher level of insulation is possible evacuating the perlite, with a resulting thermal conductivity up to 40 times less than 0.029 W/m/K depending on vacuum and temperature. It also may be used for storage of oxygen, nitrogen, and LNG when especially low thermal conductivities are desired. At the present stage we believe that such an improvement is not necessary, since additional losses are anyway due to the circulation of the liquid and the presence of cables and other ducts at cryogenic temperature.